\documentclass{aa} 
\usepackage{graphicx}

\usepackage{txfonts}
\usepackage{natbib}
\bibpunct{(}{)}{;}{a}{}{,}
\begin{document}

\title{The long bar as seen by the VVV survey: I. Colour-magnitude diagrams}

\author{C. Gonz\'{a}lez-Fern\'{a}ndez\inst{1}
          \and
          M. L\'opez-Corredoira\inst{2,3}
          \and
          E. B. Am\^ores\inst{4,5}
          \and
          D. Minniti\inst{6,7}
           \and
          P. Lucas\inst{8}
           \and
          I. Toledo\inst{6}
          }
\institute{Departamento de F\'{\i}sica, Ingenier\'{\i}a de Sistemas y Teor\'{\i}a de la Se\~{n}al, Universidad de Alicante, Apdo. 99, E03080 Alicante, Spain\\
              \email{carlos.gonzalez@ua.es}
            \and
            Instituto de Astrof\'{\i}sica de Canarias, E-38200 La Laguna, Tenerife, Spain
            \and
            Departamento de Astrof\'{\i}sica, Universidad de La Laguna, E-38206 La Laguna, Tenerife, Spain
            \and
            FCUL, Campo Grande, Edificio C5, 1749 - 016 Lisboa, Portugal
            \and
            Laborat\'{o}rio Nacional de Astrof\'{i}sica MCTI, Rua Estados Unidos 154, Itajub\'{a}$-$MG, 37504-364, Brazil
            \and
            Departamento de Astronom\'{\i}a y Astrof\'{\i}sica, Pontificia Universidad Cat\'olica de Chile, Casilla 306, Santiago, Chile
            \and
            Vatican Observatory, V00120 Vatican City State, Italy
            \and
            Centre for Astrophysics Research, University of Hertfordshire, College Lane, Hatfield AL10 9AB
            }

   \date{Received ---; accepted ---}
\abstract
   {The VISTA Variable Survey (VVV) is able to map the Galaxy at $l\leq 0^\circ$ with an unpaired depth (at least 3 mag deeper than 2MASS), opening new possibilities for studying the inner structure of the Milky Way.}
   {In this paper we concentrate on the exploitation of these data to better understand the spatial disposition and distribution of the structures present in the inner Milky Way, particularly the Long Bar and its interaction with the inner disc.}
   {To attain this, we present the Ks vs J-Ks diagrams obtained with VVV of regions with ($-20^\circ <l<-8^\circ$, $|b|<2^\circ $). From them we derive the distribution of red clump giants with heliocentric distance as a proxy for the overall stellar structure of the Galaxy. Along with these diagrams, we analysed the distribution of photometrically selected red supergiants, in order to detect events of recent stellar formation.}
   {The observations show the presence of a clear overdensity of stars with associated recent stellar formation that we interpret as the traces of the Long Bar, and we derive an angle for it of $41^\circ\pm5^\circ$ with line Sun-Galactic centre, touching the disc near $l=27^\circ$ and $l=-12^\circ$. The colour-magnitude diagrams presented here also show  a lack of disc stars in several lines of sight, a fact that we associate with the truncation of the disc by the potential of this bar for $R_{\mathrm{GC}}\leq5~\mathrm{kpc}$.}
   {}

	\titlerunning{The long bar as seen by VVV}
	\maketitle

\section{Introduction}

The signs that the Milky Way is a barred galaxy have increased following the many discoveries about the morphology of the central regions of the Galaxy, most coming from detailed near-infrared (NIR) star counts. A bar-like distribution in the stars of our Galaxy is observed from the asymmetries in the infrared surface brightness maps \citep[e. g.][]{Dw95} and in source counts \citep[e.g. ][]{St94, H94}, which both systematically show more stars at positive Galactic longitudes within $<30^\circ $ and close to the Galactic plane compared to the negative side. 

While some papers refer to a thick triaxial structure (bulge or thick bar) with a semi-major axis of around 2.5 kpc  and a position angle (PA) of 15$^\circ $--30$^\circ $ with respect to the Sun-Galactic centre direction \citep[e.g.][]{LC05, Ha06, Ra07, Va09}, other contributions suggest that there is also a long thin bar, the in-plane bar, with 4 kpc in semi-major axis length and a position angle of around 45$^\circ $ \citep[e.g.][]{LC07, CL07, Vl08, Ch09}. This long bar has a tip in the positive longitude at the beginning of the Scutum arm \citep{Da86}. As realized by \citet{Sev99}, the angle is around 25 deg. when low latitudes are excluded from the fit, and around $45^\circ$ in the plane regions within $-15^\circ<l< 30^\circ$.  

Rather than two structures, \citet{Mv11} suggested that there is a scenario of a single structure with a twisted major axis. We find this proposal acceptable from a purely morphological point of view. However, bars form as a result of an instability in differentially rotating discs \citep{Sel81}, whereas bulges are a primordial galactic component. Therefore, the scenario bulge+bar is not the same thing as a single structure, and one should observe other differences apart from the morphology. We think that the particular features of this new proposal of a ``single twisted bulge/bar'' scenario leaves some observational facts unexplained, whereas the model of a misaligned bulge+long bar successfully explains them all \citep{LC11}. Furthermore, there are two distinct populations in the thick bulge. The metal-rich population presents bar-like kinematics, while the metal-poor population shows kinematics corresponding to an old spheroid or a thick disc, so the two main scenarios for the bulge formation co-exist within the Milky Way bulge \citep{Ba10}.

We pay attention here to the long bar. We explore it in the negative longitudes ($-20^\circ <l<-8^\circ$, $|b|<2^\circ $) with the data of 2MASS+VISTA-VVV in the NIR. For the positive longitudes there are available studies including deep NIR surveys \citep[e.g.][]{CL07, CL08}. Four aspects are interesting to explore in this region:
\begin{enumerate}
\item The stellar population of the bar itself, which can be followed by a tracer, such as the red clump population, as already carried out in the positive Galactic longitudes \citep{H00, CL07, CL08}.
\item The tip of the long bar in the negative longitudes, which is tentatively placed at $l\approx -14^\circ $ \citep{LC01}. Finding there a massive star formation region would be new supporting evidence for the presence of a long bar in our Galaxy, as it was for the tip of the bar in the positive
longitudes \citep{LC99, N11, Gr11, NG11}. Such regions can form because of the concentrations of shocked gas where a galactic bar meets a spiral arm, as observed at the ends of the bars of face-on external galaxies.
\item The dust lane before the long bar in the negative galactic longitudes \citep{Ca96, Ma08}.
\item The stellar density of the inner disc around and behind the long bar+bulge, and particularly the possibility that the in-plane regions have a deficit of stars, due to a truncation in this density or to the effect of a flare in the disc \citep{LC04}. 
\end{enumerate}
All these features will be analysed using NIR colour-magnitude diagrams of the stellar populations, deriving the extinction, and separating within them the red clump stars (old population) and the red supergiants or very bright red giants (young population).

This paper is linked to Am\^ores et al. (2012), where the stellar counts produced with the same datasets used here in the form of CMDs are analysed. After a brief introduction of the datasets that will be used, a detailed description of the analysis tools we used follows in Sect. \ref{dataa}. In Sect. \ref{results} the main results of the paper are given, leading to a summary of the more relevant conclusions in Sect. \ref{discu}. 

\section{Data}

\subsection{VISTA-VVV}
The ESO public near-IR variability survey VVV is scanning the Milky Way bulge and an adjacent section of the mid-plane with high star formation activity \citep{Mi10}. The survey will take 1929 hours of observations with the four metre VISTA telescope, across an area of approximately $550~$square degrees in the inner galaxy. The final products will be a deep IR atlas in five passbands ($ZYJHK_\mathrm{S}$) and multi-epoch photometry in the $K_\mathrm{S}$-band to build a catalogue of about a billion point sources, including about a million variable sources. The main goal of the VVV survey is to enable the construction of a 3-D map of the inner Milky Way using well-understood distance indicators, such as RR Lyrae, Cepheids and red clump giants.

The first dataset, including observations from year 2010, has just been released. \citet{Sa12} discuss the observations, reductions, photometry, calibrations, and catalogues. For most regions of interest the limiting magnitude is $m_{\mathrm{K_S}}\sim18.0$, and the image scale is $0.34~\mathrm{"/pixel}$. A more detailed description of the VVV data used can be found in Am\^ores et al. (2012).

\subsection{2MASS}
Although we are interested in the inner Milky Way, best viewed with VVV, we also study brighter populations, some of which might be too bright for this survey. We supplement the VVV data using 2MASS All-Sky Release \citep{St06}, accessing its point source catalogue through the Gator interface\footnote{http://irsa.ipac.caltech.edu/applications/Gator/}.

\section{Data analysis}
\label{dataa}
\subsection{Combining 2MASS and VVV}

We have made intensive use of the colour-magnitude diagram (CMD) instead of a source-by-source analysis, so a detailed combination of both datasets is not needed. Instead, for a more subtle approach we opted to adopt hard thresholds in $K_\mathrm{S}$ and $J$: above these thresholds we rely on 2MASS photometry, and below them VVV is the data source.

Determining these thresholds requires careful calculation of the completeness of 2MASS. We restricted ourselves to $-20^\circ<l<-8^\circ$, and in these regions completeness is dominated by source confusion, owing to the high stellar density. For each of the fields we built a histogram in magnitude, fit the brighter part with a second degree polynomial, and checked when the observed counts fall below $80\%$ of the predicted value offered by this fit, adopting this value as our completeness limit. As we want to have a homogeneous set of CMDs, we selected a single set of thresholds for all our fields: $(J, K_\mathrm{S})=(14,12.5)$.

As we are comparing data from two different surveys, with different instrumental configurations and calibration procedures, we also checked that both photometries are consistent. As both catalogues overlap in magnitude, for each field we crossmatched the common stars and checked for differences in magnitude. The typical values are around $0.05~\mathrm{mag}$ in $J$ and $0.02~\mathrm{mag}$ in $K_\mathrm{S}$, although in some fields subject to severe reddening these values can be higher, reaching differences up to $0.4~\mathrm{mag}$. In all our subsequent analysis, these systematic differences have been added to VVV magnitudes.

We also checked the behaviour of the magnitude difference against colour and found no evident dependence, but in the redder part of the CMD (with $J-K_\mathrm{S}>4.5$), there are few common stars, so this effect could be masked.

\subsection{The infrared colour magnitude diagram}

\begin{figure*}
\includegraphics[width=6.0cm,angle=90]{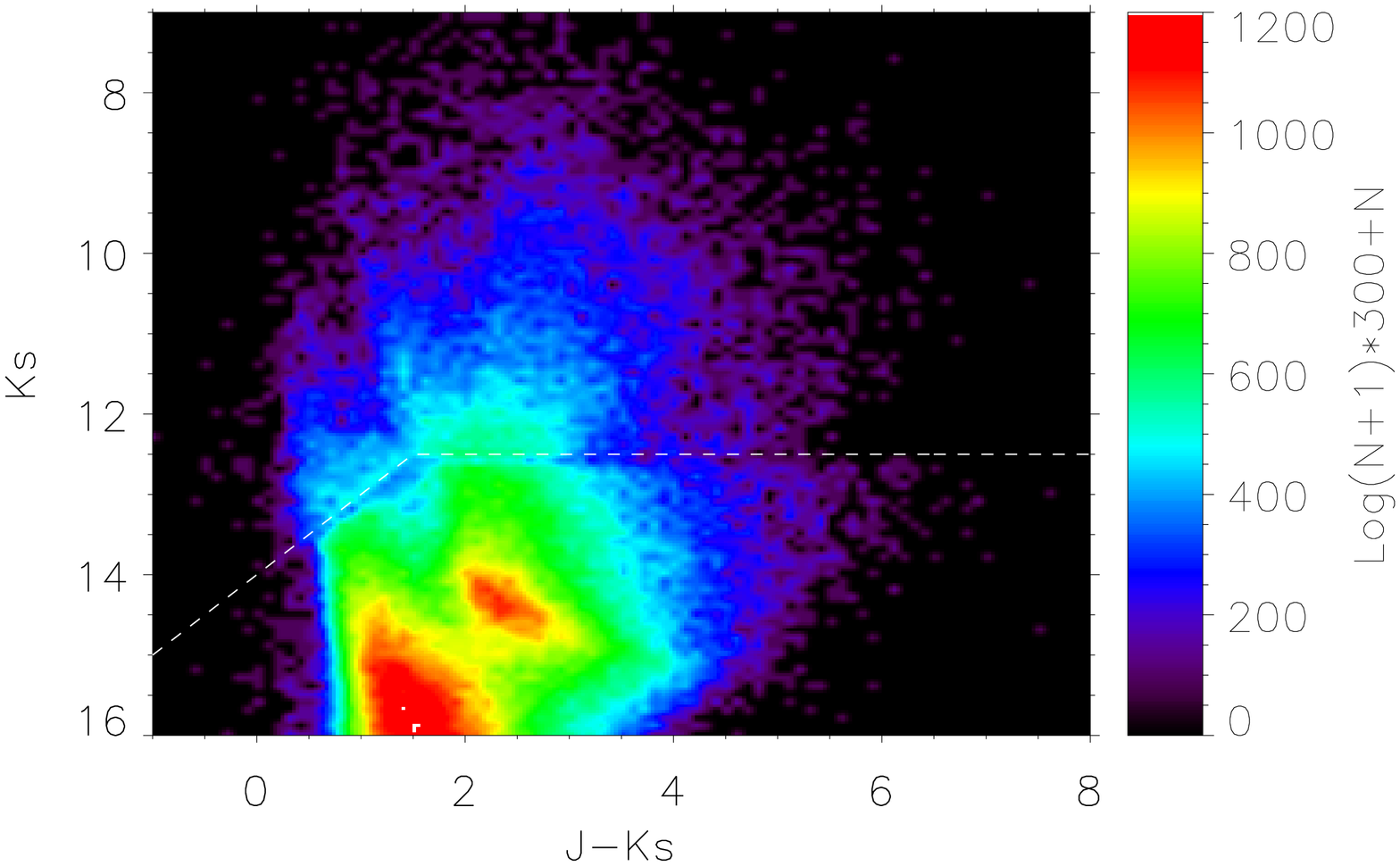}
\includegraphics[width=8.0cm]{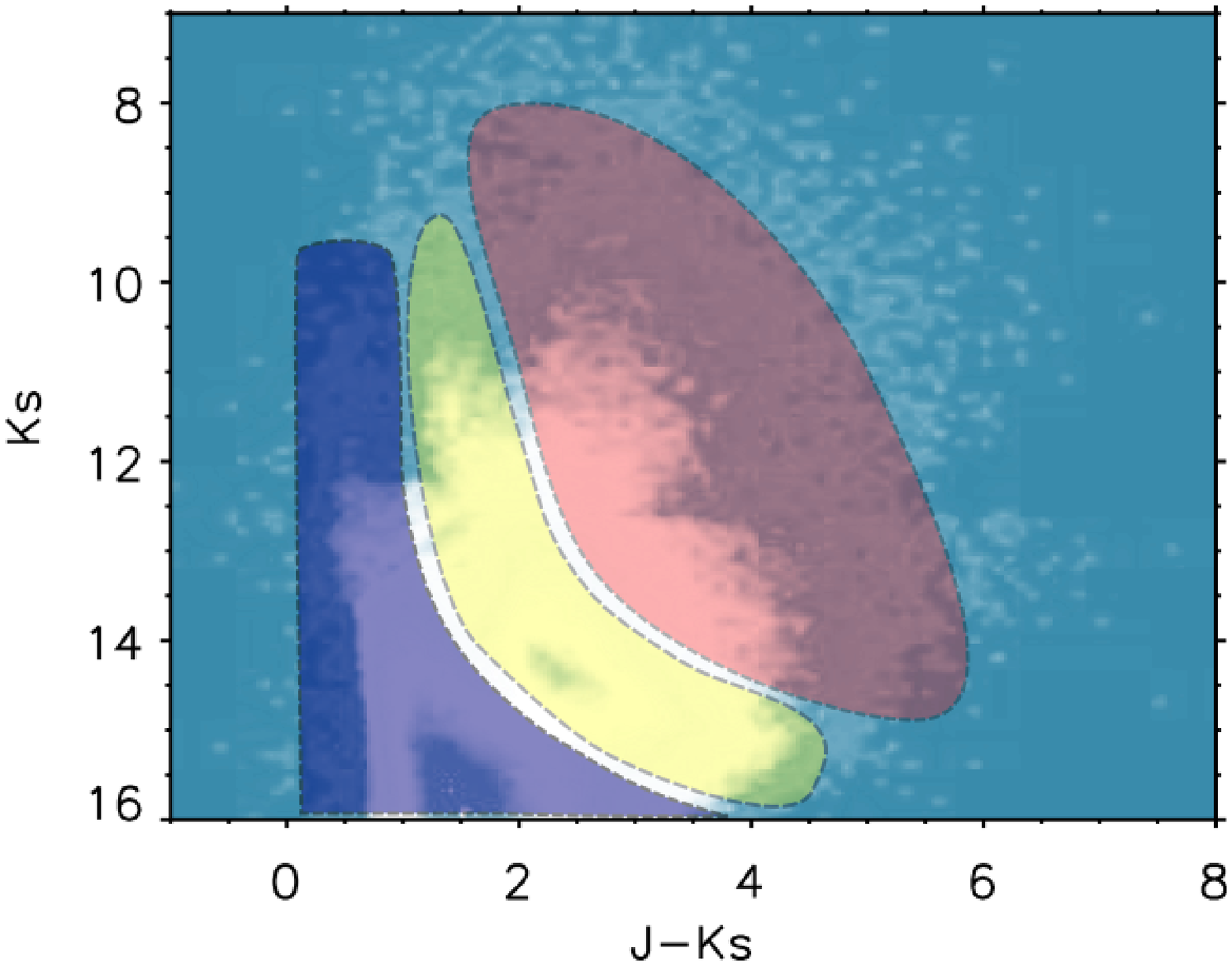}
\caption{Left: Sample CMD of a $0.5^\circ\times0.5^\circ$ field around $(l,b)=(-11^\circ,0^\circ)$, combining data from VVV and 2MASS. The white dashed line marks the transition between both surveys. Right: Sketch of the same plot with the regions that occupy dwarfs (blue), RCGs (yellow), and later-type giants (red), as discussed in the text.}
\label{DCM}
\end{figure*}

Since the advent of large-scale infrared surveys, the infrared CMD, usually of $(J-K_{\mathrm{S}})$ versus $K_{\mathrm{S}}$ has become one of the standard tools for studying the structure of the Milky Way. Beyond the penetrating powers of this wavelength range that allow us to reach the inner parts of our Galaxy, in these diagrams it is relatively easy to separate the two main visible populations, dwarfs and giants. Although the morphology of the CMD changes greatly in different areas of the Galaxy, since they are heavily affected by interstellar extinction, in-plane fields offer a similar aspect, exemplified in Fig. \ref{DCM}.

This CMD was obtained for a field centred at $(l,b)=(-11^\circ,0^\circ)$. The main features that dominate the figure are outlined in the right-hand panel: in blue  and occupying the lower left-hand region of the plot, the dwarfs of the disc are the more abundant population, separated from the giants due to the extinction and their intrinsic brightness. Even if caused by distance, dwarfs and giants could appear to us with the same apparent magnitudes, since giants are intrinsically brighter they would have to be farther away and be the subject of heavier interstellar extinction, since we are dealing with in-plane fields. This implies that they will appear shifted towards the red, occupying then the top right-hand part of the diagram. 

The effect of distance (that moves stars towards dimmer apparent magnitudes) and extinction (that makes stars appear dimmer and redder) results in an ideal population of stars with the same intrinsic magnitude and colour and evenly distributed along the line-of-sight that would appear as a diagonal line (not necessarily straight) in this CMD. The natural spread in magnitude of any real population changes this line into a strip of a certain width, and the uneven stellar density of the Galaxy translates into overdensities in the $(J-K_{\mathrm{S}},K_{\mathrm{S}})$ space. Because the later the type of a giant the brighter and redder it becomes, the lower envelope of this region is composed by late type G giants, but the bulk of it will correspond to giants of spectral types around K2, also known as red clump giants (RCG), which are the most abundant giant population. The redder tail of Fig. \ref{DCM} would be composed almost exclusively of M giants. The relative abundance of RCGs is high enough so that they appear as a distinct strip on the CMD while later type giants, much less abundant, are visible towards redder and brighter magnitudes.

To build the CMDs used in this work (including the one from Fig. \ref{DCM}), we always use $0.5^\circ\times0.5^\circ$ fields centered on the nominal coordinates. As the source density is high, instead of plotting star by star, we built a density diagram using tiles of $0.1~mag$ in colour and magnitude, as can be seen in Fig. \ref{DCM}. Also, the variation range of $N$, the number of stars in each one of these tiles, is wide, so that to appreciate all the substructure present in these CMD it is mandatory to plot instead some function that compresses this range. After a lengthy trial and error process we opted for the function
\begin{equation}
I=300\times\log{(N+1)}+N,
\end{equation}
where $N$ is the number of stars per tile and $I$ the plotted intensity. This selection combines the compression factor of a logarithmic intensity plot but without losing the ability to show density variations of $\sim100$ stars per tile, changes that in a pure logarithmic plot would be invisible. As our analysis of these density variations was purely qualitative, we opted to use a function that maximizes the visible information instead of one with immediate physical translation, such as a square root or a logarithm.

The resulting CMDs, used in subsequent sections, can be seen in Appendix \ref{appdcm}.

\begin{figure*}
\includegraphics[width=6.6cm,angle=90]{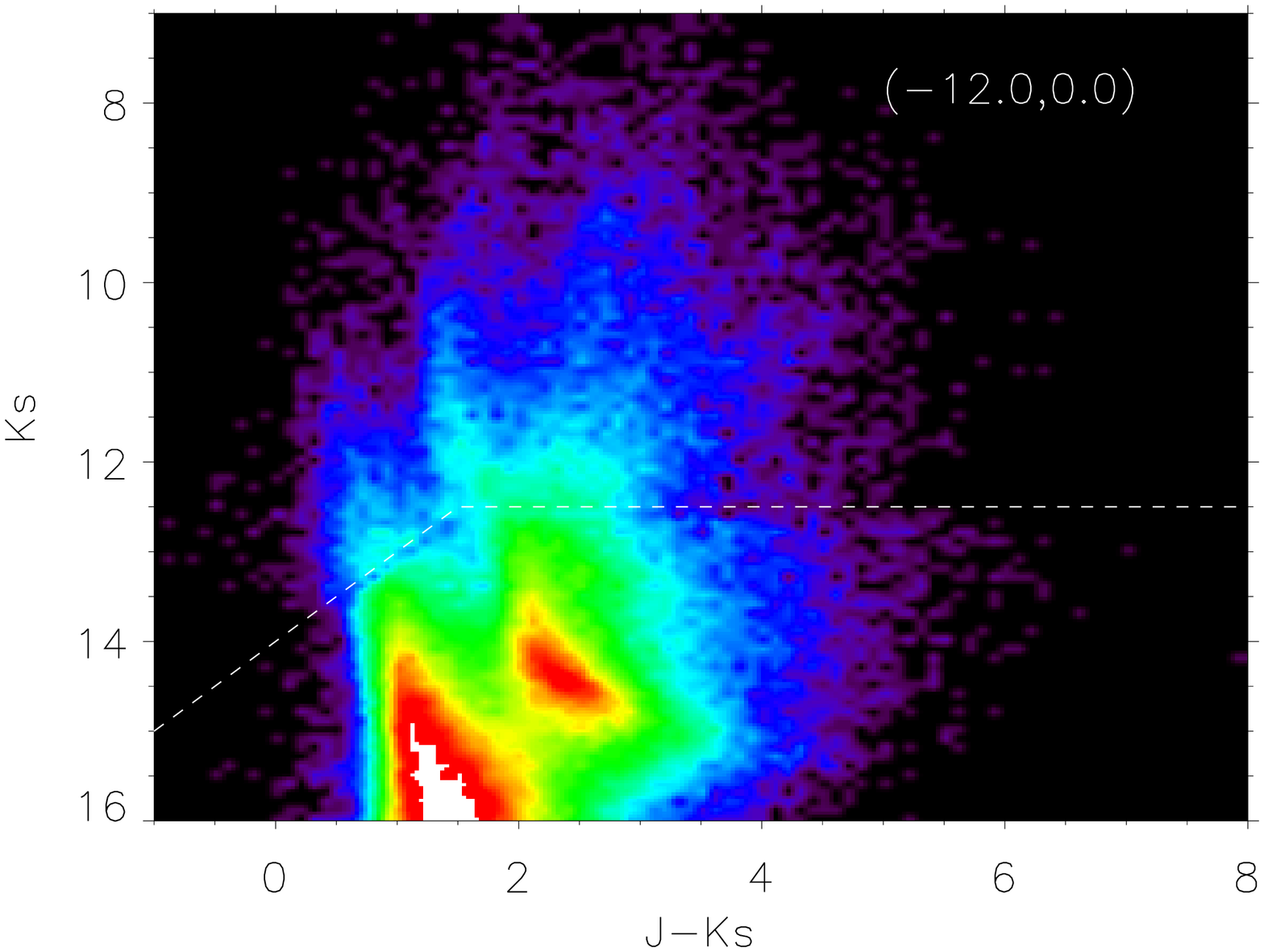}
\includegraphics[width=6.6cm,angle=90]{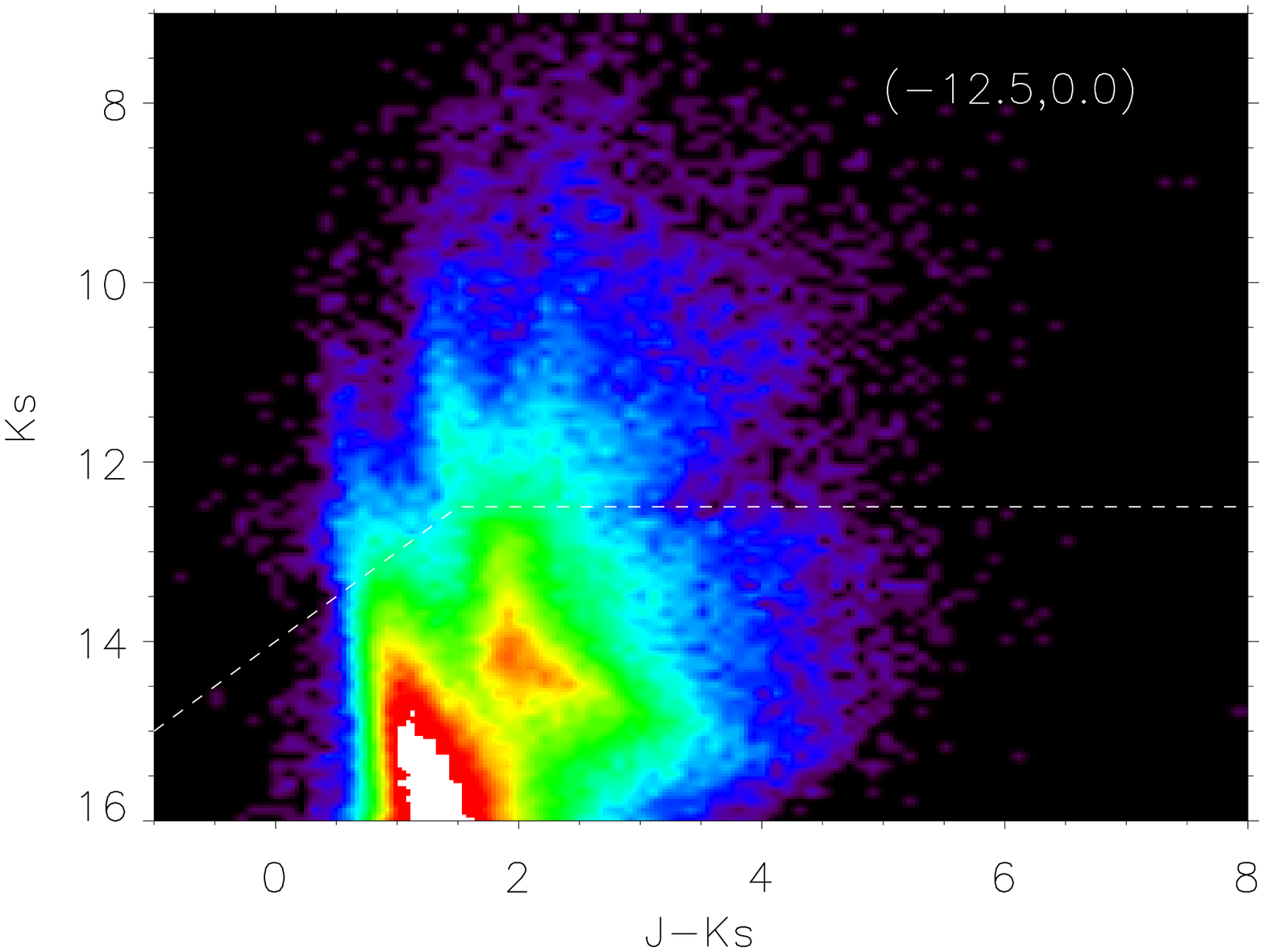}\\
\includegraphics[width=6.6cm,angle=90]{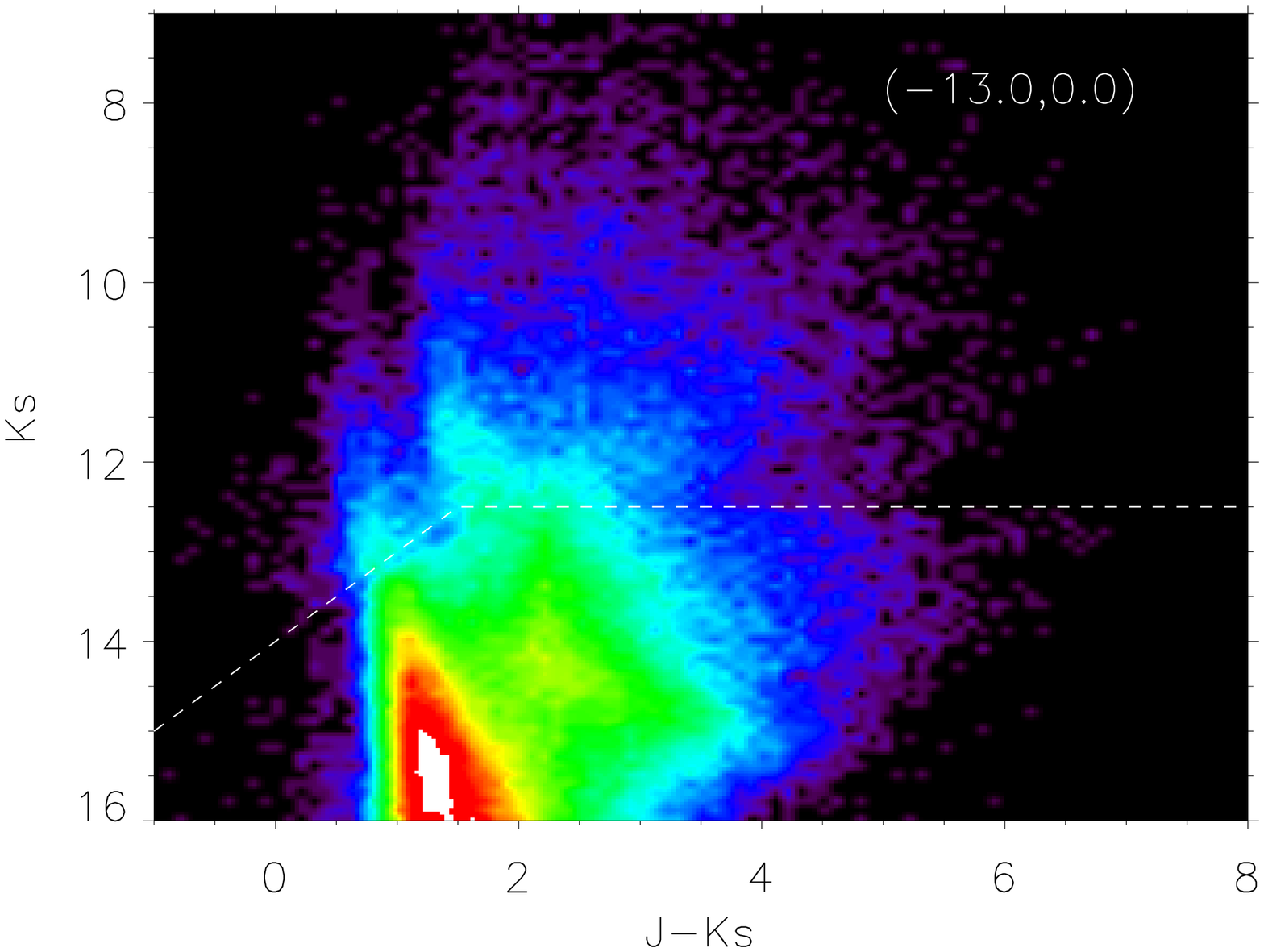}
\includegraphics[width=6.6cm,angle=90]{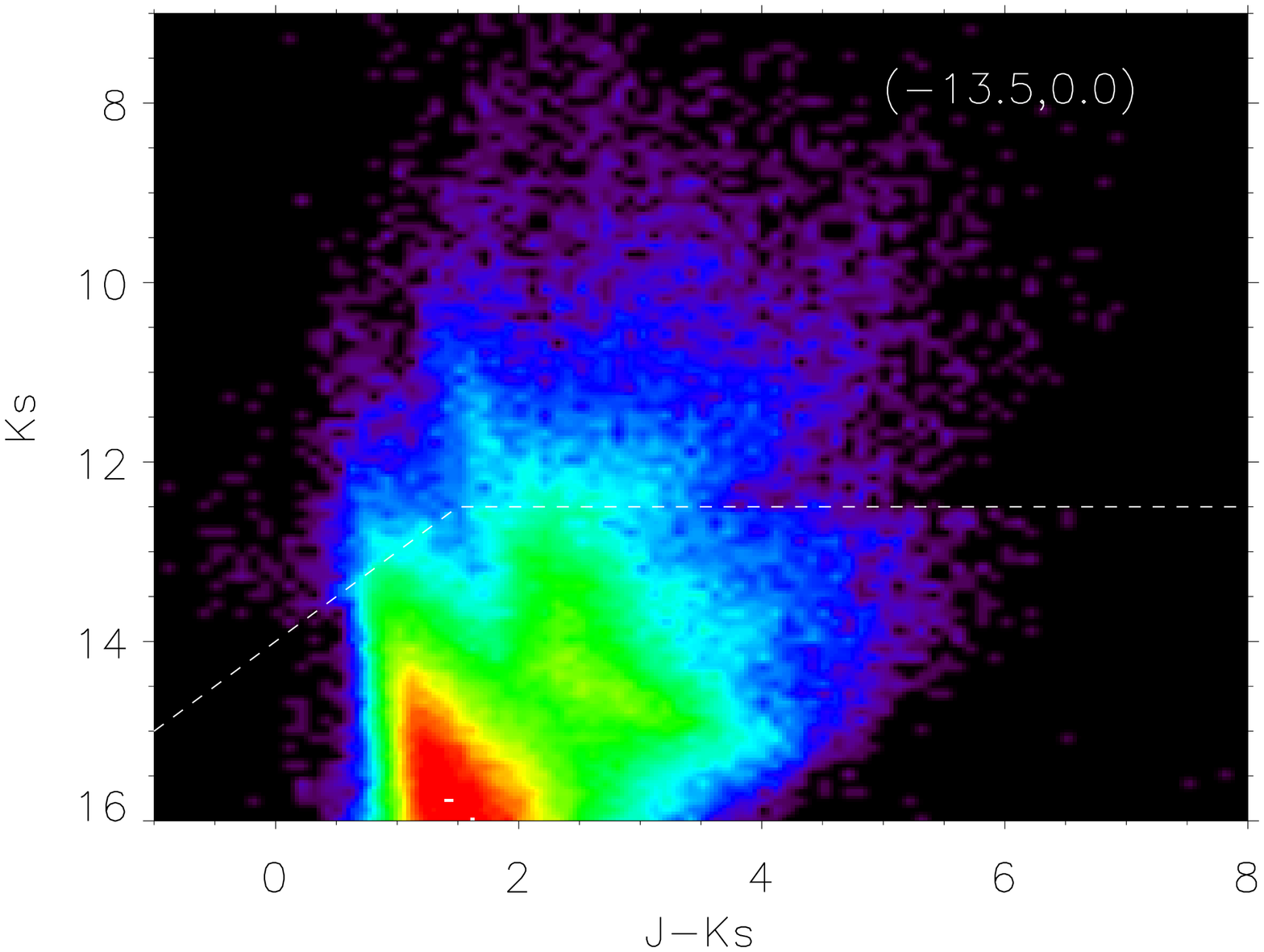}
\caption{CMDs of the in plane fields ($b=0^\circ$) between $l=-12.0^\circ$ and $l=-13.5^\circ$. The colour scale is the same as Fig. \ref{DCM}, and the white dashed line represents the transition from 2MASS to VVV data. The rather sudden drop in the number of stars between $l=-12.5^\circ$ and $l=-13.0^\circ$ is evident in this plot.}
\label{starfall}
\end{figure*}

\subsection{The red clump method}
\label{RCGm}

As said above, the ubiquity of the RCGs makes the so-called red clump method one of the standard tools for analysing CMDs in the infrared. Its use and application is discussed in depth in \citet{LC02} (hereafter LC02), but in this section we offer a brief outline of its application to the 2MASS+VVV composite CMDs, as the results of this method is used thoroughly through the text.

These RCGs have a very narrow luminosity function that it is also calibrated well (LC02), so their intrinsic colour and luminosity are well known. It is possible then to transform their position in the $(J-K_{\mathrm{S}},K_{\mathrm{S}})$ plane to $(d,A_{\mathrm{K}})$,  once we assume an extinction law (in the form of the relative extinction coefficient $A_{\mathrm{J}}/A_{\mathrm{K}}$). A more detailed description of the method and the errors involved can be found in \citet{CL08} and references therein. Following this study, we use $(J-K_{\mathrm{S}})_0=0.70\pm0.05$, $M_{\mathrm{K}}=-1.62\pm0.03$ and $A_{\mathrm{J}}/A_{\mathrm{K}}=2.518$.

To perform this calculation there is a previous step that needs to be taken: the calculation of the $(J-K_{\mathrm{S}},K_{\mathrm{S}})$ curve that represents the distribution of RCGs over the CMD. The simplest way to do this is to cut the CMD into several strips in $K_\mathrm{S}$ and then fit the $(J-K_{\mathrm{S}})$ distribution of each strip with the sum of a second degree polynomial and a Gaussian; the $\mu$ of this gaussian gives the fiducial point that represents the locus of the RCG for that strip (LC02). The problem here is that, in the presence of high interstellar extinction (see for example the CMDs for $b=0^\circ$ in Fig. \ref{av2}), the RCG strip almost appears horizontal, and so projecting the data of each strip over the $(J-K_{\mathrm{S}})$ is not optimal, because it mixes stars with very different colours but with roughly the same $K_{\mathrm{S}}$ magnitude, resulting in a loss of information.

To overcome this, as a first step in our analysis, a rough fit of the RCG strip is performed by eye, and this $(J-K_{\mathrm{S}},K_{\mathrm{S}})_0$ curve is used as the entry point of an automated process that, instead of performing cuts over $K_{\mathrm{S}}$ and projecting them over the  $(J-K_{\mathrm{S}})$ axis, defines a local coordinate system $(X,Y)$ that is perpendicular to this $(J-K_{\mathrm{S}},K_{\mathrm{S}})_0$ curve at each point. As these new axes take the reddening vector into account, stars with similar magnitudes but different reddening appear separated when projecting over either $X$ or $Y$, so a much more refined  $(J-K_{\mathrm{S}},K_{\mathrm{S}})$ curve can be derived.

Once we have the variation in $(J-K_{\mathrm{S}})$ with $K_{\mathrm{S}}$ for the RCGs, we can extract this population from the CMD. Although this can be optimized for each line-of-sight, and because our study requires the comparison of data at different Galactic longitudes, we consider that everything within $0.5~\mathrm{mag}$ in colour and magnitude from this curve is a RCG.

\subsection{Photometric search for massive stars}
\label{rsgsearch}

One of the best tracers of stellar formation regions are red supergiants (RSG). Since these stars are very bright in the infrared \citep[with a conservative limit of $M_\mathrm{K}<-8$,][]{W83}, they shine through the obscuration present in the plane. One of the expected effects of a galactic bar is a triggering of star formation at its tips and, depending on the bar strength, over its dust lanes \citep[see, for example,][]{Sh02}. As such, the study of the distribution of RSGs with Galactic longitude could give us clues on the structure of the bar.

In the near end of the bar, around $l\sim28^\circ$ several clusters of these stars have already been detected \citep[i.e.][]{N11, CGF12}. In these clusters, RSGs will have apparent magnitudes $7<m_\mathrm{k}<4$. If we assume that the bar forms a $45^\circ$ angle with the Sun-Galactic centre line and has a $4~\mathrm{kpc}$ semimajor axis \citep{CL08}, this would put the far end of the bar at $\sim11.2~kpc$, while the near one sits at $\sim5.9~kpc$. This difference, in distance modulus, is of $\sim1.4~\mathrm{mag}$, so the same RSG that we see at the near end of the bar would have magnitudes in the range $9<m_\mathrm{k}<6$, allowing for a $\sim0.5~\mathrm{mag}$ in extra extinction due to the presence of the dust lane of the bar.

It is worth noting that, although we have selected a $\phi=45^\circ$ bar, other possible position angles have been postulated \citep[for example,][with $\phi=22^\circ\pm5.5^\circ$]{Ba05}, but always with $\phi<45^\circ$. This implies that for these bar configurations, the far end should be farther away than for a $45^\circ$ angle bar. The difference in brightness for a given RSG between both configurations is only of $0.1~\mathrm{mag}$ (corresponding to $0.6~\mathrm{kpc}$), and our selection criteria for RSG is therefore not biased against these models.

For this reason, the possible RSGs situated at the far end of the bar would be very difficult to separate from the population of late type giants (they also have very similar intrinsic colours) of the disc. One possible way to separate this population is to use the reddening free pseudo-colour $Q$, defined as $Q=(J-H)-1.8\times(H-K_\mathrm{S})$ \citep{N07}. Generally speaking, the later (and intrinsically redder) a star is, the higher value of $Q$ it will have, with late type giants (types beyond K0) having $Q>0.41$. RSGs, because their atmospheres are very extended, complex, and often present mass loss, will not deredden properly, and so they roughly occupy the range $0.1<Q<0.4$ \citep{Me11}.

Another applicable threshold affects the colour of the candidates. RSGs are intrinsically red sources, with intrinsic colours $(J-K_\mathrm{S})_0>0.7$. Since we are also looking at sources under heavy extinction (as is expected for the inner Milky Way), we can safely assume that any RSG will have $(J-K_\mathrm{S})>1.3$. This step eliminates late type dwarfs from the disc that will fulfill most of the previous criteria but, as they are located on the local disc will have lower colour excesses.

Summing up, the criteria that we used to select RSG candidates over 2MASS data are
\begin{enumerate}
\item $0.1\leq Q\leq0.4$,
\item $m_\mathrm{K}\leq9$,
\item $(J-K_\mathrm{S})\geq1.3$.
\end{enumerate}

\citet{N11} performed a spectroscopic survey around the massive cluster RSGC3, confirming the nature of stars selected based on these criteria, and find that the initial sample is entirely composed of RSGs and very evolved giants, with stellar types later than M5. A $38\%$ fraction of the objects turned out to be class I sources (supergiants), while $28\%$ were class II (bright giants), and the remaining $34\%$ identified as late class III giants.

\section{Results}
\label{results}
\subsection{Structure of the Galactic bar}
\label{density}
One of the most striking facts when analysing the CMDs produced with 2MASS+VVV for fields with $|l|<0^\circ$ and $b=0^\circ$ is that there is a sudden drop in the number of giants around $l=-13^\circ$ (see Fig. \ref{starfall}). Since the total number of sources per square degree would be dominated by the dwarf population of the disc and we are more interested in the innermost region of the Milky Way, we focus our study on the giants, that reach beyond the disc within the limiting magnitude of VVV.

\begin{figure}
\resizebox{\hsize}{!}{\includegraphics[angle=90]{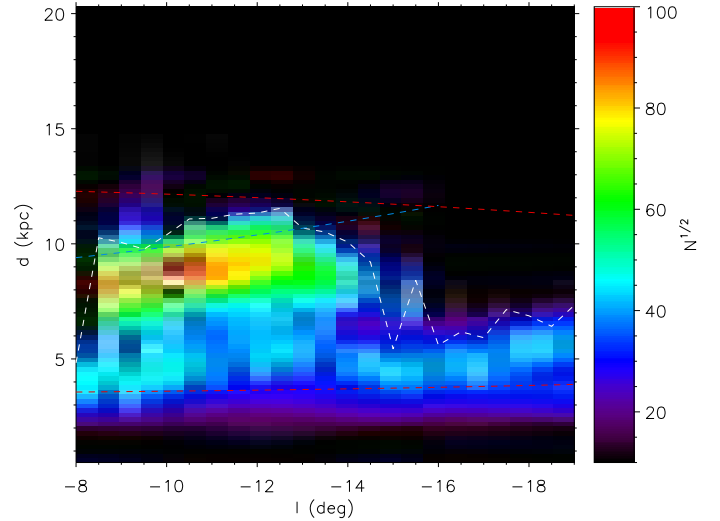}}
\caption{Number of red giants per square degree as a function of Galactic longitude and distance for fields with $b=0^\circ$. To increase contrast, the plotted quantity is the square root of this number, $N^{1/2}$. The white dashed line represents the completeness limit at each line-of-sight. The blue and red dashed lines mark the place of a $45^\circ$ bar and a ring of $r=4.5~\mathrm{kpc}$, respectively. The sudden drop at $l=-8^\circ$ is due to the combination of higher reddening and stellar density. All these giants verify $|b|\leq0.5^\circ$.}
\label{ngiants}
\end{figure}

\begin{figure}
\resizebox{\hsize}{!}{\includegraphics[angle=90]{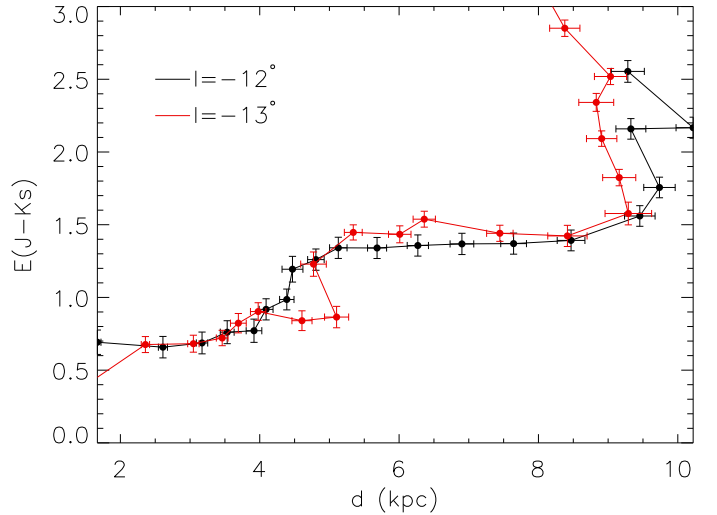}}
\caption{Variation in $E(J-K_\mathrm{s})$ with the distance along the line-of-sight for the in-plane fields at $l=-12^\circ$ (black) and $l=-13^\circ$ (red). The strange behaviour seen in this last curve at $\sim4.5~\mathrm{kpc}$ is due to the imperfect combination of 2MASS and VVV data in this line-of-sight. For an explanation of the negative slope in parts of the curves, see Sect. \ref{appext}.}
\label{l12ext}
\end{figure}

Using the method described in Sect. \ref{RCGm}, we can extract the RCGs from a CMD and estimate their distance. This allows us to reconstruct the in-plane 2-D distribution of these stars, as can be seen in fig. \ref{ngiants}. Beyond $l=-8^\circ$, where extinction and source density affect our sample, the behaviour of the number of giants with $(l,d)$ is rather smooth, showing traces of a rather straight structure that ends beyond $l\sim-12^\circ$, where there is an appreciable drop in the number of visible giants. This could easily be associated with a bar-like structure, but before doing this there are some issues that need to be tackled. This drop in the number of giants visible in Fig. \ref{starfall} could be due to an increment in the extinction along the line-of-sight, but this is unlikely: in this same figure it can be seen that the morphology of the RCG strip is similar in all the lines of sight, with higher latitudes and lower extinction. This implies a similar $(d,A_\mathrm{K})$ behaviour for all of them. Using the method described in Sect. \ref{RCGm}, we can estimate the variation of $E(J-K_\mathrm{S})$ with $d$, plotted in Fig. \ref{l12ext}. Comparing it with Fig. \ref{starfall} we see that, although the extinction along the line-of-sight is almost equal for both of them, the number of visible giants is appreciably lower for $l=-13^\circ$.

\begin{figure}
\resizebox{\hsize}{!}{\includegraphics[angle=90]{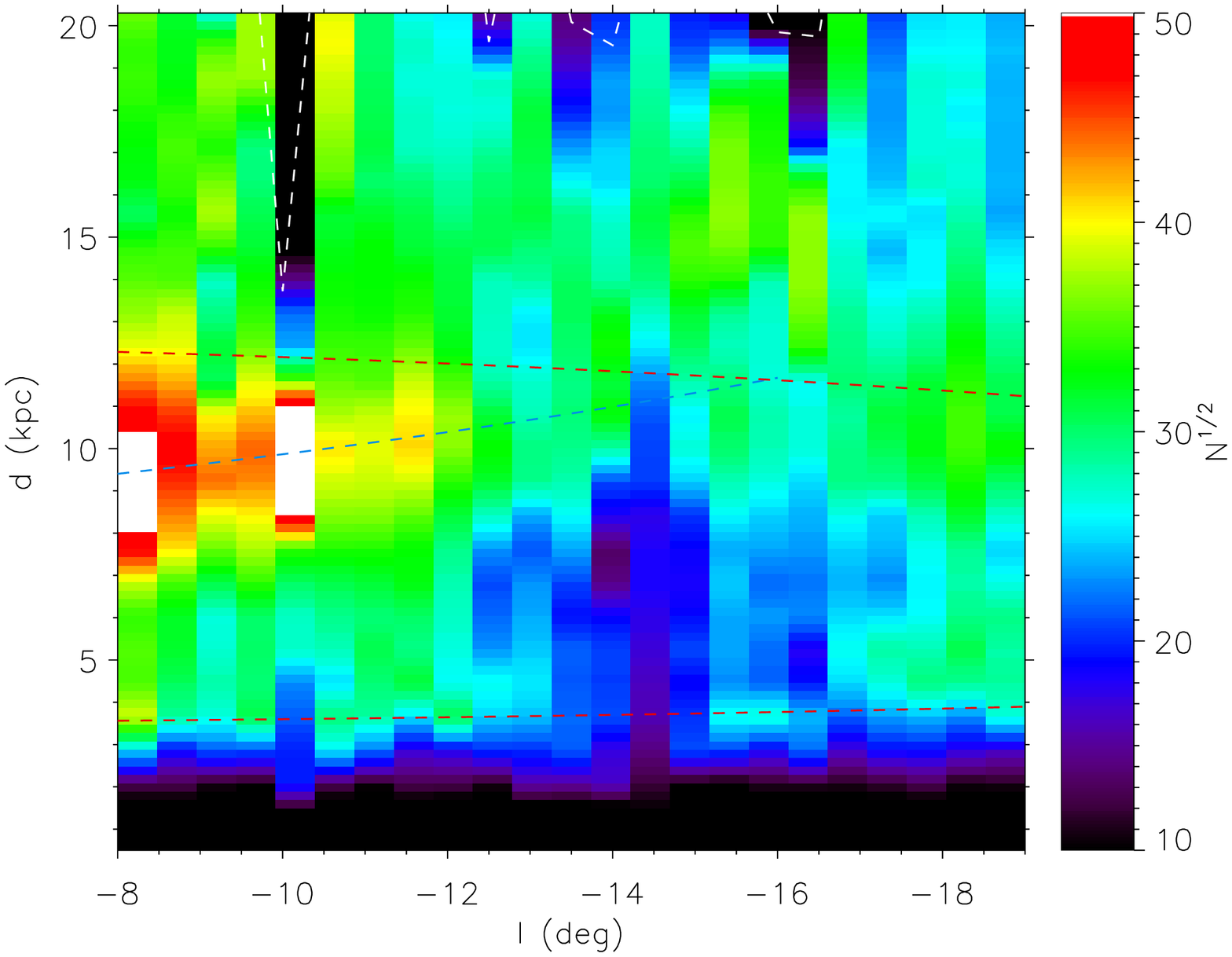}}
\resizebox{\hsize}{!}{\includegraphics[angle=90]{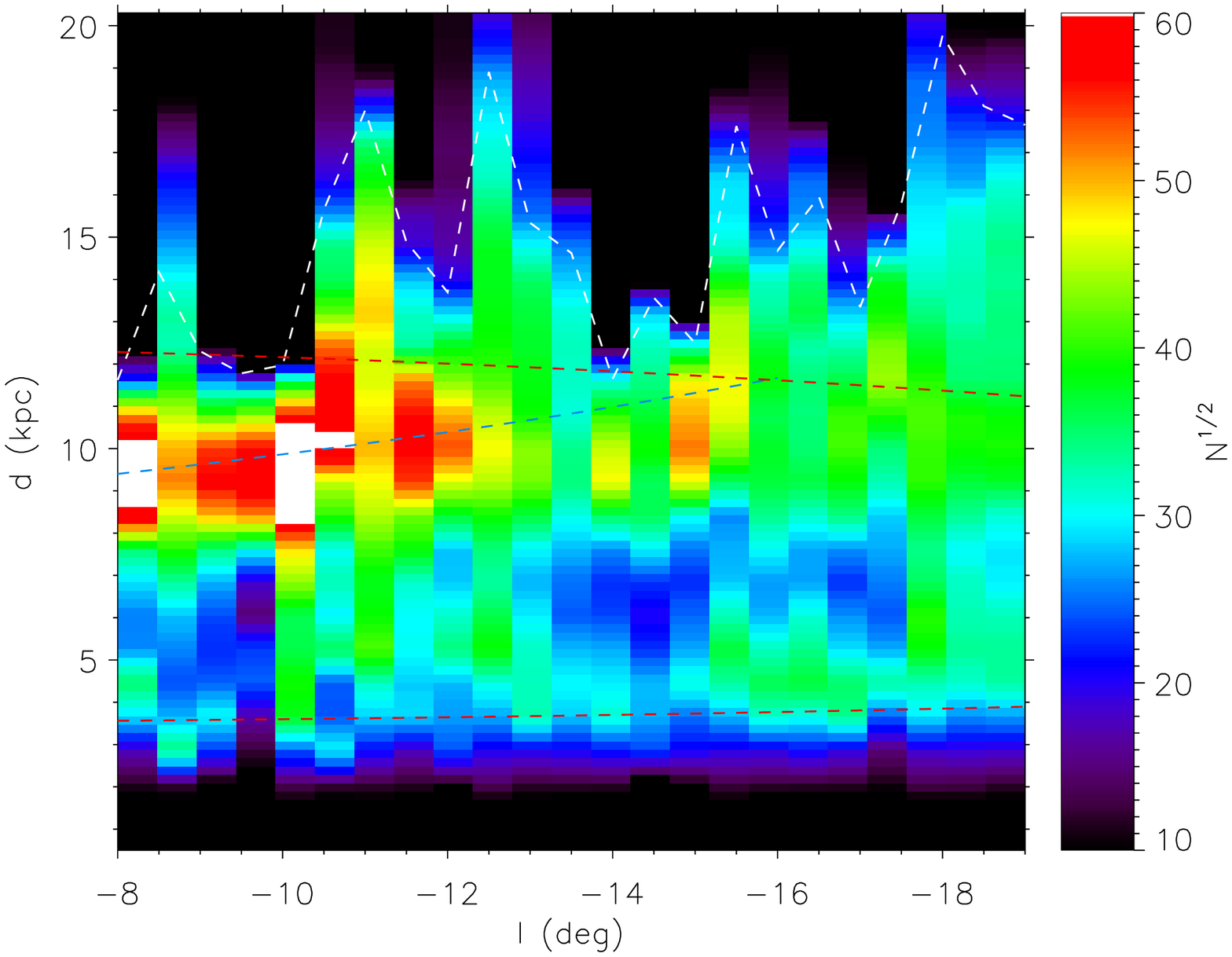}}
\resizebox{\hsize}{!}{\includegraphics[angle=90]{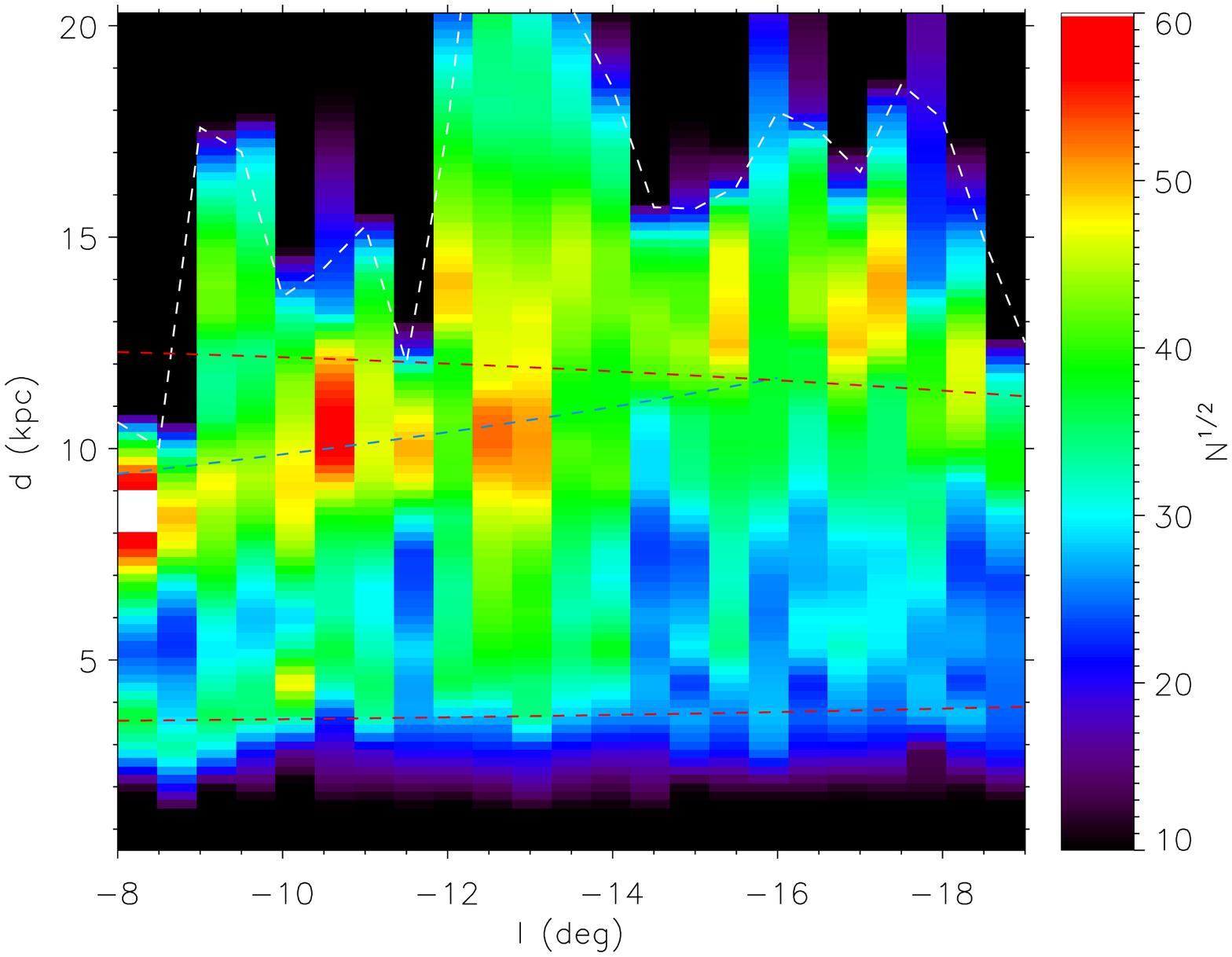}}
\caption{Same plot as Fig. \ref{ngiants}, but for RCGs with  $-0.75^\circ\geq b \geq -1.25^\circ$, (bottom), $0.75^\circ\leq b \leq1.25^\circ$ (middle), and $1.25^\circ\leq b \leq1.75^\circ$ (top). Notice the difference in scales within these plots and also with Fig. \ref{ngiants}}
\label{ngiants15}
\end{figure}

The density distribution of Fig. \ref{ngiants} could also be an effect of the variations in the limiting magnitude with Galactic longitude. A simple exponential density distribution, when observed with the instrumental setup of VVV, would show a maximum in the number of stars near completeness, for when we approach this limit we start to gradually lose more visible stars, modulating the subjacent stellar distribution. As hinted before, in off-plane fields these effects are less severe, and in Fig. \ref{ngiants15} we can see how this bar-like structure is visible, well before the completeness limit is reached.

This structure corresponds very well to the extension of the Long Bar, as has been observed previously in integrated light \citep{H00} and stellar counts \citep{LC01}, to the fourth quadrant. Assuming that this bar is symmetric with respect to the Galactic centre, the positions of its tips provide a strong constraint on its size and orientation. We place its far end at $l=-13^\circ$ (as suggested by Figs. \ref{ngiants}, \ref{RSGplane}, and \ref{ngiants15}), and the nearby end in $l=27^\circ$ \citep{CL08}. Although these measurements are not very precise, the error should not exceed $\pm1^\circ$. Assuming that $R_{\mathrm{GC}}=8.0\pm0.5~\mathrm{kpc}$, we obtain an angle of $41^\circ\pm5^\circ$ and a total length of $8.2\pm0.6~\mathrm{kpc}$. This is close to the values calculated by \citet{LC01, CL07, CL08}, but clearly different from the proposed geometry of \citet{Ba05}.

The structure in Figs. \ref{ngiants} and \ref{ngiants15} seems to have a wider angle, close to $45^\circ$ but the range in galactic longitudes is small, and these differences in orientation are difficult to judge. Also, the derivation of $d$ for this figure is affected by extinction, which could warp the visible structure if there is variation between different lines of sight. The properties derived using only the angular positions of the bar ends are very robust, not affected by any of these effects, the distance to the Galactic centre being the only assumption made.

\begin{figure}
\includegraphics[width=6.6cm,angle=90]{l120b00.eps}
\includegraphics[width=6.6cm,angle=90]{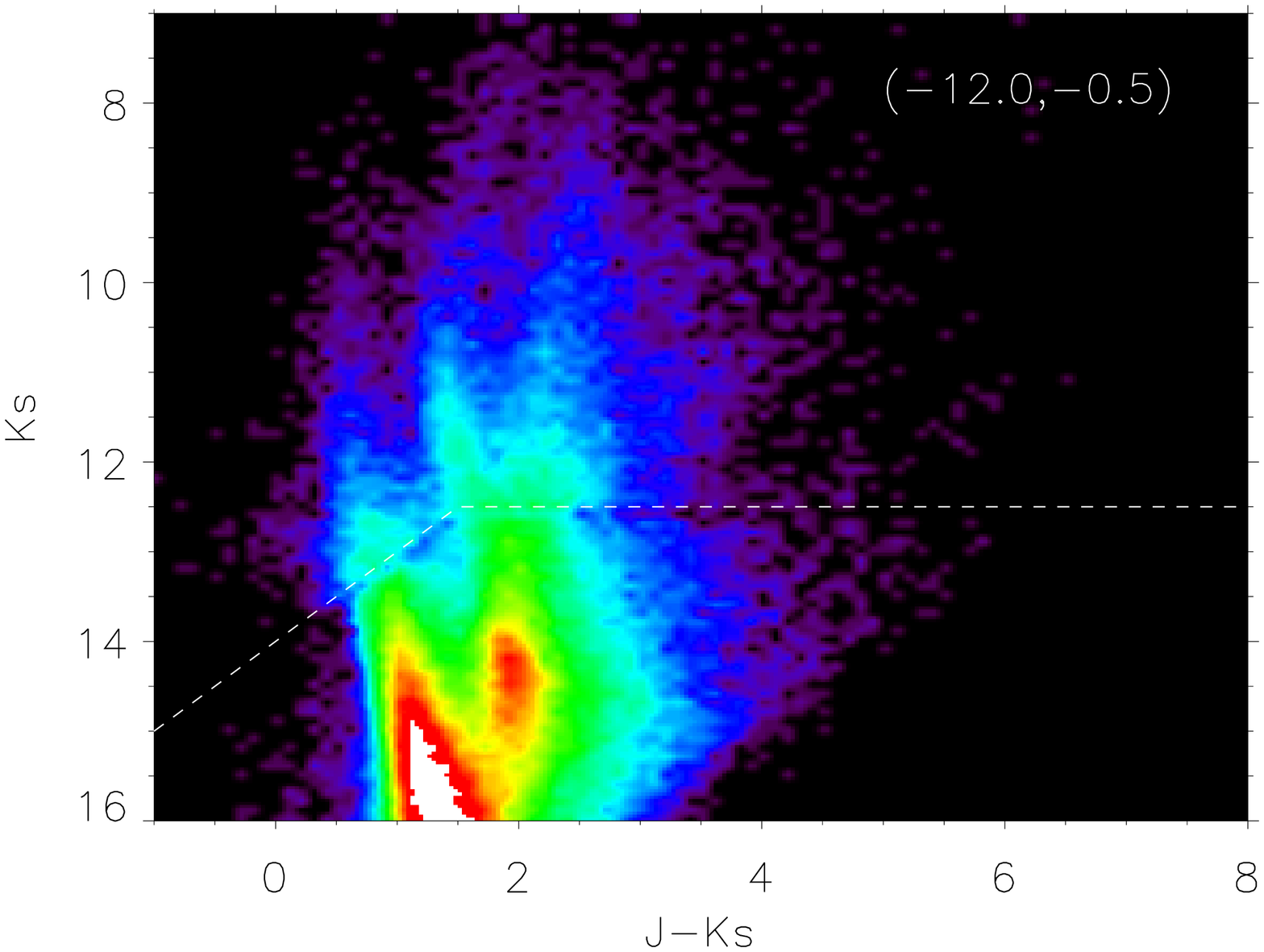}
\includegraphics[width=6.6cm,angle=90]{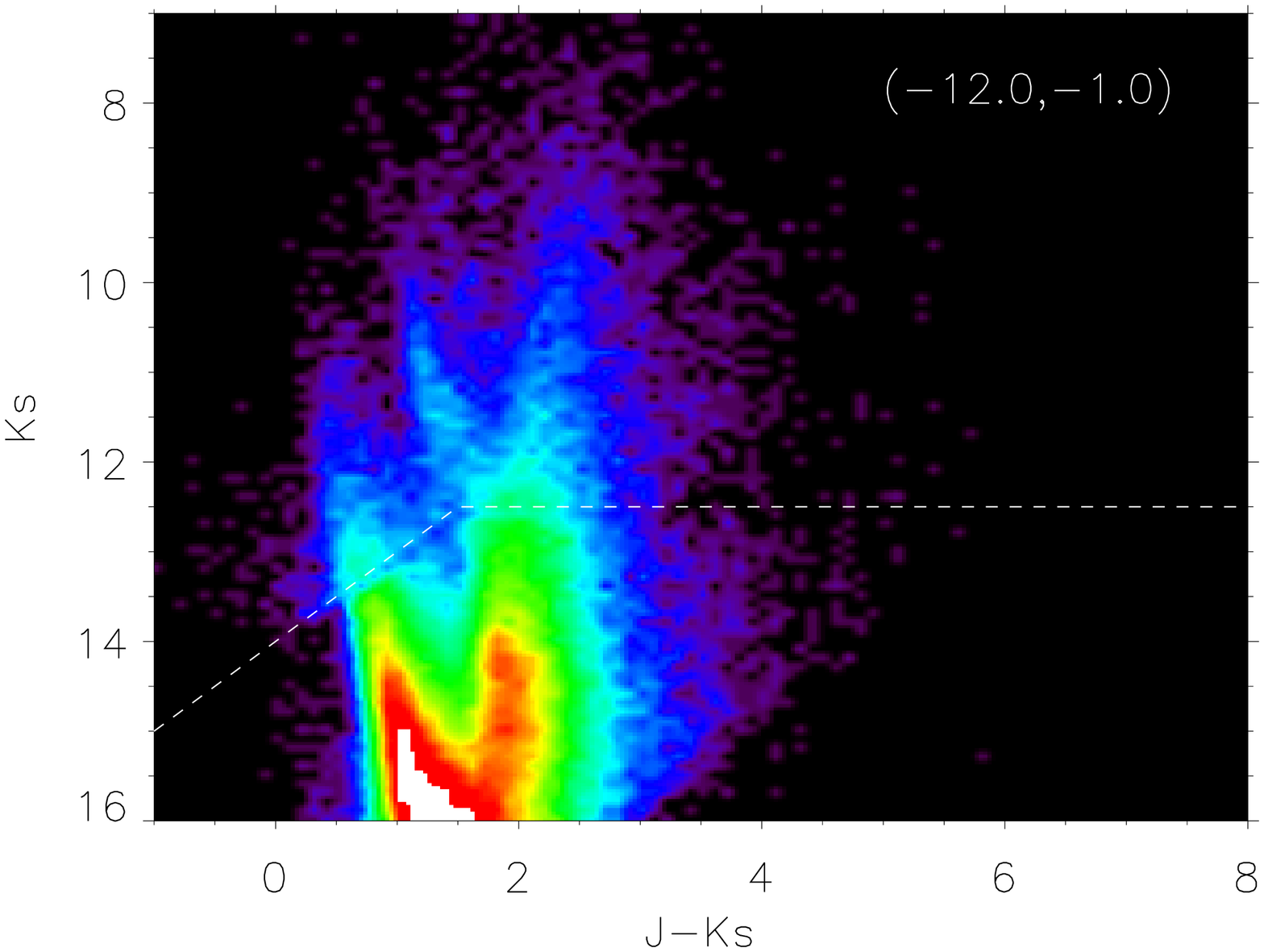}
\caption{CMDs of the fields with $l=-12.0^\circ$ between $b=-1.0^\circ$ and $b=0.0^\circ$. The colour scale is the same as Fig. \ref{DCM}. As can be seen, there is a sudden increase in the extinction at $(J-K_\mathrm{S},m_\mathrm{K})\sim(2,14)$ in the plane that is not visible at higher $|b|$. Also, the second maxima of the RCGs is visible in the bottom plot, roughly at $(J-K_\mathrm{S},K_\mathrm{S})=(2,15)$}
\label{vstruct}
\end{figure}

Another interesting feature that, for example, can be observed in the CMDs of Fig. \ref{starfall} and the extinction curves from Fig. \ref{l12ext} is that there is a huge amount of interstellar material concentrated in a narrow strip around $9~\mathrm{kpc}$ at $l=-13^\circ$ (although this distance estimate varies with the assumed extinction law, see Sect. \ref{appext}). This wall of extinction makes the RCGs appear as an almost horizontal strip on the CMD, but its effect is constrained to the plane. In Fig. \ref{vstruct} we plot the diagrams of a vertical cut at $l=-12.5^\circ$ (near the end of the bar), and we can see clearly how the effect of interstellar extinction decreases from $b=0.0^\circ$ to $b=-0.5^\circ$. This feature of the interstellar medium can be clearly followed in the plane from $l=-13^\circ$ to $l=-8^\circ$ (see Figs. \ref{av5} to \ref{ab0}).

In these longitudes it is easy to see that this structure is too far away to be mistaken for the 3-kpc arm/molecular ring (contrary to what might happen in the first quadrant), so this sudden increase in the extinction (hence, the density of interstellar material) is a clear clue to the existence of a leading dust lane in the Long Bar of the Milky Way. A similar structure was previously detected in the inner Milky Way over extinction maps and molecular emission \citep{Ma08}, although at lower $|l|$. Dust lanes are a common companion to stellar bars in other galaxies \citep[for example,][]{Co09}, so the presence of such a structure in our Galaxy is a hint of the existence of a bar up until $l\sim-12^\circ$.

\begin{figure}
\resizebox{\hsize}{!}{\includegraphics[angle=90]{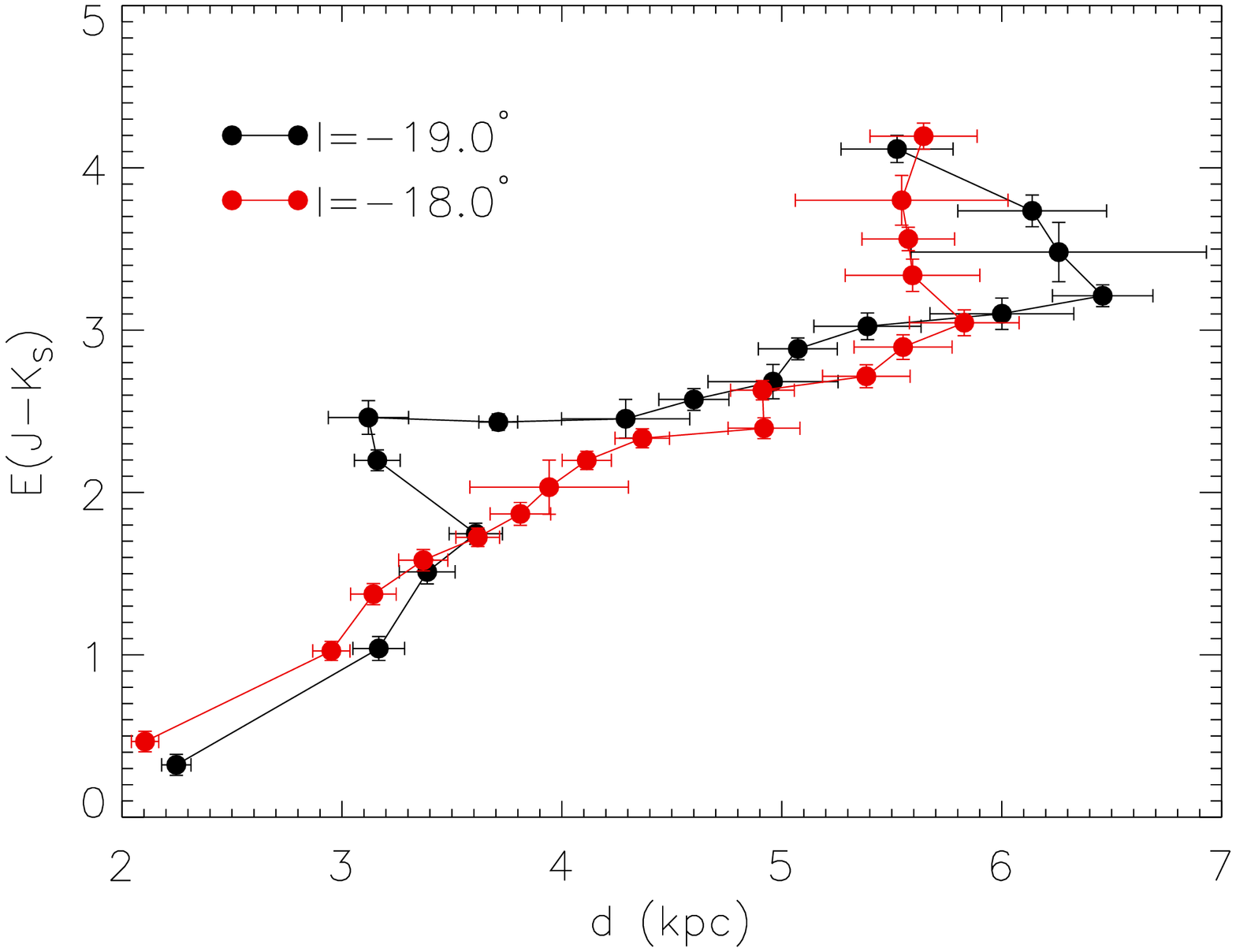}}
\caption{Same as Fig. \ref{l12ext} but for the in-plane fields at $l=-18^\circ$ and $l=-19^\circ$.}
\label{ext2}
\end{figure}

As we move outwards, this dust lane disappears, but the interstellar extinction grows more or less continuosly for $l<-14.5^{\circ}$. As can be seen in Fig. \ref{ext2}, the variation in $E(J-K_\mathrm{S})$ with $d$ is quite different from that of the dust lane. Instead of a sudden increase in the amount of interstellar material, the variation of Fig. \ref{ext2} is much smoother. This matches very well what is expected of a near-tangent line-of-sight to a gas rich structure such as an arm (see Sect. \ref{drive}).

\subsection{Structure of the inner disc}

\begin{figure}
\resizebox{\hsize}{!}{\includegraphics{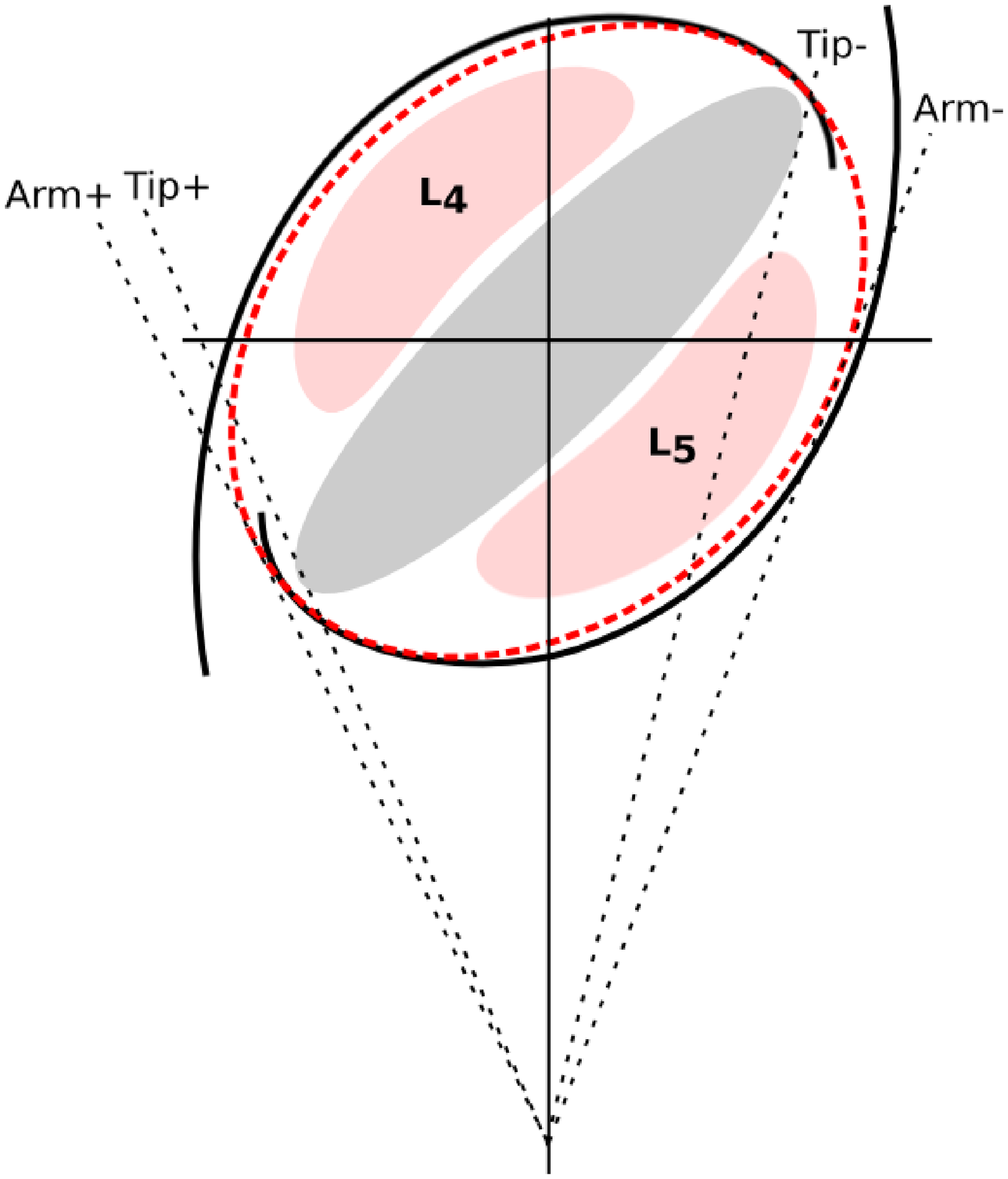}}
\caption{Schematic view of the inner Milky Way, showing two possible configurations: a ring (red dashed line) and two arms (solid black lines). While both are almost indistinguishable in the first quadrant, it is possible to distinguish between them in the far end of the bar and the tangent line to the arm. The light-red shaded areas represent the areas where the maxima of the potential of the bar ($\mathrm{L_4}$ and $\mathrm{L_5}$) sweep away stars and gas. The bulge has been left out of this representation.}
\label{schem}
\end{figure}

A careful look at Fig. \ref{ngiants15} presents us with another interesting feature: the depletion of giants present from $-18^\circ\leq l\leq -12^\circ$ (particularly evident at $b=1.5^\circ$). This truncation of the stellar density of the disc is visible in every galactic latitude except for $|b|\leq0.5^\circ$, due to completeness (see Fig. \ref{ngiants}). This is one of the strongest visible effects of a bar; its typical potential \citep{Bi08} has two maxima in the direction of its minor axis \citep[labelled $\mathrm{L_4}$ and $\mathrm{L_5}$ in Fig. \ref{schem}, following the notation of][]{Bi08}. These maxima sweep away stars and gas outwards, effectively emptying the disc inside the corotation radius. In Fig. \ref{ngiants15} this appears clearly in the first plot, and there are traces of the same effect in the third, but masked mostly by the effects of extinction. The method described in Sect. \ref{RCGm} assumes that the extinction varies smoothly along the line-of-sight, so the clumpiness of the real distribution of $A_\mathrm{K}$ with $d$ translates into a cumulative blur of the features in Figs. \ref{ngiants} and \ref{ngiants15}.

This truncation is not only visible in the stellar content, but it also affects interstellar matter. As can be seen in Fig.\ref{l12ext}, from $\sim4.5$ to $\sim9~\mathrm{kpc}$ there is little variation in $E(J-K_\mathrm{S})$ with $d$. As the colour excess is proportional to $A_\mathrm{K_S}$ and this quantity, in turn, is a function of the integral of the volume of interstellar matter along the line-of-sight, a flat curve implies that there is little to no underlying gas, matching the behaviour of the giants in Fig.\ref{ngiants15}.

Even so, the photometric depth of VVV is high enough to probe the far disc for the first time. In some of the CMDs of appendix \ref{appdcm}, the trace of the RCG runs vertically for $m_\mathrm{K}\geq12$ (particularly evident for $(l,b)=(-12.5,-1.0)$). This trace shows a secondary maximum, visible on several lines of sight (as, for example, in Fig. \ref{vstruct}). These maxima appear around $13~\mathrm{kpc}$, and are clearly visible in the reconstructed distribution of Fig. \ref{ngiants15} (more evident in the middle and bottom panels).

This increase in stellar density shows us where the disc regains its density beyond corotation. This effect has also been noted in \citet{Go11} at $|b|=1^\circ$, although the authors interpret these maxima as traces for an arm at $11.2~\mathrm{kpc}$. This seems unlikely, since a one-degree separation from the plane implies a height of $200~\mathrm{pc}$ at $11~\mathrm{kpc}$ ($300~\mathrm{pc}$ for a $1.5^\circ$ line-of-sight), and arms are structures enclosed in the thin disc, so have very small height-scales. On top of this, the effective volume of an arm that a line-of-sight crossing it probes is very small unless at a tangent point. Since these structures are mostly composed of young populations, they are not traced very well with RCGs (a late population, with $\tau\sim2~\mathrm{Gyr})$, and it is very unlikely that they could appear clearly at any line-of-sight, as in Fig. \ref{ngiants15}.

\subsection{The behaviour of interstellar extinction}
\label{appext}

Many of the previous results depend, or at least are affected by, extinction. In the derivation of Figs. \ref{ngiants} and \ref{ngiants15} we have assumed a simple extinction law that remains constant with $d$, $l$, and $b$ and that yields $A_\mathrm{J}/A_\mathrm{K}=2.518$. As can be seen in Fig. \ref{l12ext}, beyond $9~\mathrm{kpc}$, the slope of the $(d,E(J-K_\mathrm{S}))$ curve becomes negative. This is physically impossible, and is a direct effect of this choice of extinction law, for the derivation of $d$ depends on the calculation of $A_\mathrm{K}$ and, as follows:
\begin{eqnarray*}
E(J-K_\mathrm{S})=(J-K_\mathrm{S})-(J-K_\mathrm{S})_0\\
E(J-K_\mathrm{S})=A_\mathrm{J}-A_\mathrm{K_S}=A_\mathrm{K_S}\left(\frac{A_\mathrm{J}}{A_\mathrm{K_S}}-1\right)
\end{eqnarray*}
This implies:
\begin{eqnarray*}
A_\mathrm{K_S}=\frac{(J-K_\mathrm{S})-(J-K_\mathrm{S})_0}{A_\mathrm{J}/A_\mathrm{K_S}-1}.
\end{eqnarray*}

Once we isolate a given population of stars that we will use as standard candles, from two measurables, $(J-K_\mathrm{S})$ and $m_\mathrm{K_S}$, we therefore derive $d$, $E(J-K_\mathrm{S})$, and $A_\mathrm{K_S}$. To complete the recipe we need to know, apart from the aforementioned $A_\mathrm{J}/A_\mathrm{K_S}$, the luminosity function of the population under study, through its $(J-K_\mathrm{S})_0$ and $M_\mathrm{K_S}$. While a variation in any of these three constants would change Fig. \ref{l12ext}, we know that:
\begin{itemize}
\item[-] An offset in $(J-K_\mathrm{S})_0$ or $M_\mathrm{K_S}$ would, in exchange, offset the curves in Fig. \ref{l12ext}, but not change their slope or overall shape.
\item[-] For the case of the RCGs, there is a known dependency of $(J-K_\mathrm{S})_0$ with metallicity \citep{CL07}, but we can see in Fig. \ref{l12ext} that, since the extinction increases in such a quick manner, most of the variation (particularly the part in which the curve acquires negative slope) takes place in less than a kiloparsec, so the metallicity spread is expected to be low. For this reason, while the overall shape of the $d$ vs. $E(J-K_\mathrm{S})$ curve might change due to this effect, this variation would affect the slope of the nearby part of the graph with respect to the points that are more distant from the observer, but it will not change the slope measured for $d>8~\mathrm{kpc}$.
\end{itemize}

Taking these points into consideration, the only remaining possibility to explain this negative slope is our choice of a value of $A_\mathrm{J}/A_\mathrm{K_S}$ not representative of the observed physical conditions. We can tinker with this value in order to bring the curves in Fig. \ref{l12ext} into coherence with the expected behaviour of extinction with distance (that, by definition, needs to be a monotonically increasing function), obtaining a lower limit for the observed value of this ratio. We find that
\begin{eqnarray*}
\frac{A_\mathrm{J}}{A_\mathrm{K_S}}\geq2.9,
\end{eqnarray*}
so far away from the typical values of \citet{RL85} or \citet{C89}, both close to $A_\mathrm{J}/A_\mathrm{K_S}=2.5$. This points to an interstellar material that is more transparent towards the red than predicted by the extinction law, more in concordance with the results from studies as \citet{Ni09} and \citet{St09}. This effect is not exclusive to the lines of sight used to produce Fig. \ref{l12ext}; heavy extinction transforms the RCG strip in a CMD into a nearly horizontal band, so similar behaviors are observed in all the fields with $-8.0^\circ<l<-13^\circ$ and $b=0.0^\circ$ in Figs. \ref{av5} to \ref{ab0}. 

A change in the $A_\mathrm{J}/A_\mathrm{K_S}$ extinction ratio implies an expansion/contraction in the distance scale in Figs. \ref{ngiants} and \ref{ngiants15}, but will not change the overall shape of the structural features discussed. Because the analysis of these figures and the results drawn from are  qualitative and the physical characteristics of the bar are calculated from the Galactic longitudes of its ends, a measure that is independent of extinction, our results are not affected by the choice of extinction law. Since a proper determination of the value of $A_\mathrm{J}/A_\mathrm{K_S}$ is beyond the scope of this work, we opt to retain the value from \citet{RL85} for homogeneity with similar studies \citep[for example,][]{CL07}

This distance scale factor introduced through  $A_\mathrm{J}/A_\mathrm{K_S}$ can be appreciated when comparing Fig. \ref{ngiants} with Fig. \ref{ngiants15}: the bar overdensity seems to shift away from the observer at higher $|b|$. As increasing $A_\mathrm{J}/A_\mathrm{K_S}$ implies decreasing $A_\mathrm{K_S}$ and a higher derived $d$, the error in this last parameter due to the difference in $A_\mathrm{J}/A_\mathrm{K_S}$ will be a monotonically increasing function of $E(J-K_\mathrm{S})$, hence higher for in-plane fields.

\subsection{Bar-driven stellar formation in the Milky Way}
\label{drive}

A structure such as the Long Bar is expected to affect its environment in several ways. One of them is the triggering of star formation where its tips interact with the Galactic disc. Since RSGs are massive stars, their lifespans are expected to be short, just a few tens of Myr, so they trace recent stellar formation.  We built a sample of photometrically selected RSG candidates following the methods discussed in Sect. \ref{rsgsearch}; namely, we selected stars with $0.1\leq Q\leq0.4$, $(J-K_\mathrm{S})>1.3$ and $m_\mathrm{K}<9$. Since these values are well within the completeness of 2MASS, we use this survey. Doing so lets us obtain candidates for all the inner Galaxy, from $l=40^\circ$ to $l=-30^\circ$.

The vertical distribution of RSG candidates is, as expected for a young population, concentrated in low $b$ fields, and it can be roughly approximated with a Gaussian of  $\sigma=2.4^\circ$. We find that in the region with $-30^\circ\leq l\leq40^\circ$ there are close to $30\,000$ candidates with $|b|\leq1^\circ$ (roughly a density of $200~\mathrm{deg^{-2}}$) and $60\,000$ in off-plane fields. 

\begin{figure}
\resizebox{\hsize}{!}{\includegraphics[angle=90]{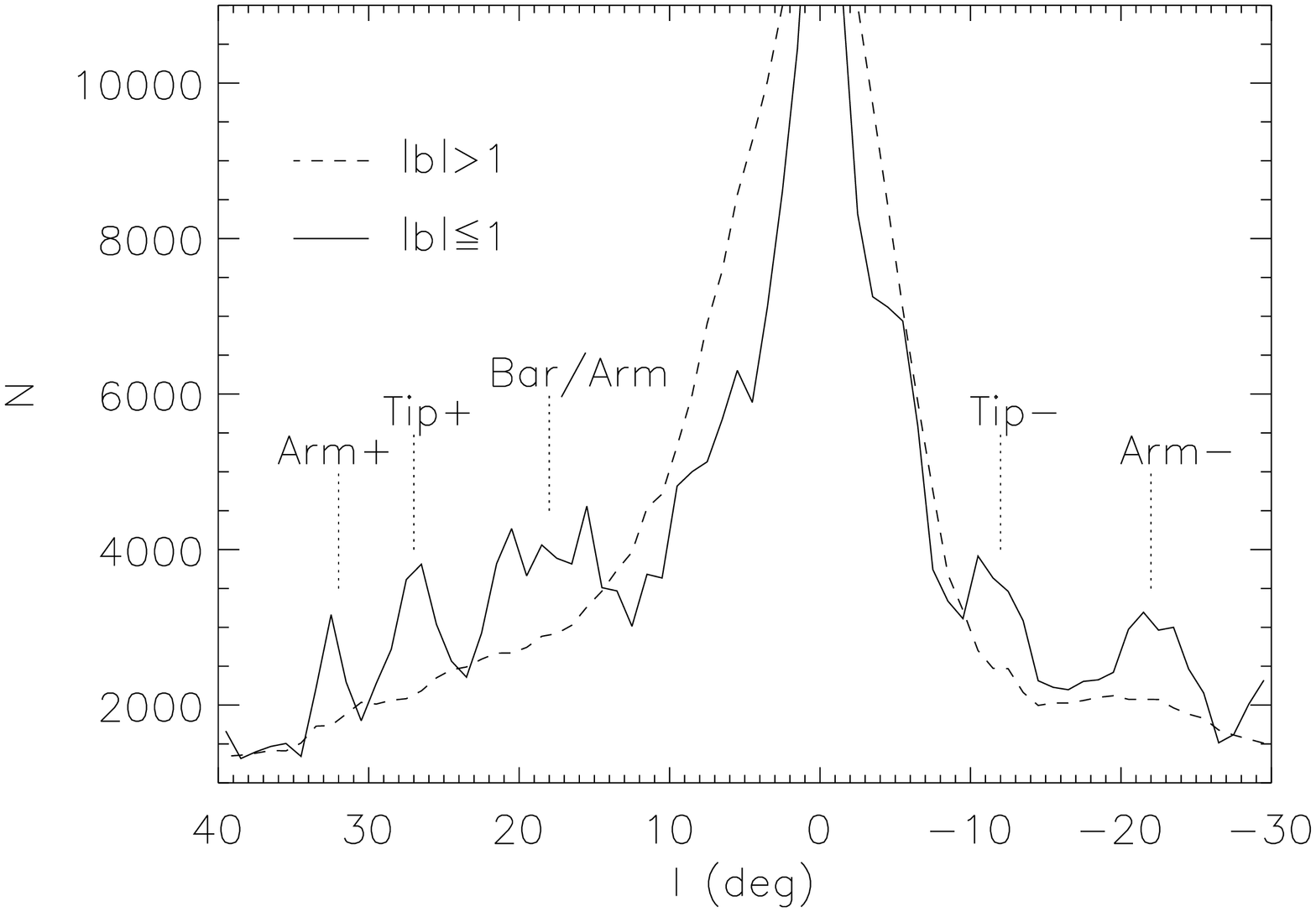}}
\caption{Number of candidate RSGs detected in 2MASS, both for fields in the Galactic plane ($|b|\leq1^\circ$) and outside ($1^\circ<|b|\leq90^\circ$). To enable easy comparison, both measurements have been normalized to the same total number of stars.}
\label{RSGplane}
\end{figure}

As reasoned in \citet{CL07}, the long bar is a very flat structure, so to locate traces of its associated stellar formation it is natural to restrict ourselves to the Galactic plane. Since several structures coexist in the region with $|l|<30^\circ$, we compared the in-plane ($|b|\leq1^\circ$) distribution of candidate RSGs with off-plane ($1^\circ<|b|\leq90^\circ$) fields. The results can be seen in Fig. \ref{RSGplane} (to facilitate comparison, the off-plane distribution has been normalized to have the same total number of objects as for in-plane fields). As can be seen, rather than being smooth, the in-plane distribution presents a series of overdensities not present  in off-plane fields (or present in a much less conspicuous manner).

These overdensities (outside the Galactic center) give us hints to possible underlying structures. This distribution follows the $2.2~\mathrm{\mu m}$ and $12~\mathrm{\mu m}$ emission studied by \citet{H94} closely. As the structure of the first quadrant of the Galaxy is much better studied, the nature of some of these peaks has already been established: the overdensity at $l=32^\circ$ is related to the tangent line-of-sight towards the Scutum-Crux arm, and its counterpart in the fourth quadrant was already established by \citet{H94} beyond $l<-29^\circ$ (hence outside our analysis). The next peak at $l=28^\circ$ is known to be related to one of the more massive episodes of star formation in our Galaxy, containing several massive clusters with $M_\odot>10^4~\mathrm{M_\odot}$, and probably with a combined mass well in excess of $10^5~\mathrm{M_\odot}$ \citep{CGF12, N11}. Since the wings of this peak are wide, it also contains the extremely massive clusters RSGC1 \citep{F06, NGF10} and Stephenson 2 \citep{D07}. All these clusters have been related to the interaction of the Long Bar and the Scutum-Crux Arm/molecular ring; this area also contains the more massive star forming region of the Milky Way, W43 \citep{NG11} (although it does not harbour any RSG yet). 

It is expected that a similar wealth of structures should occur at the far end of the bar, here identified at $l=-13^\circ$, although the associated stellar formation has yet to be discovered. Some recent studies of massive clusters have started this search, but so far the only confirmed massive association detected is Mercer 81 \citep{D12}. It is associated by the authors to the far tip of the bar, but sitting at $l=-21.6^\circ$ is likely to be part of the 3-kpc arm, with a tangent at $l=-22^\circ$ (LC01). This is a prominent feature in gas emission that was the first indication of our Galaxy being barred \citep{DeV64}. Although it was previously suspected that this arm has not been subject to recent stellar formation \citep{Da08}, in Fig. \ref{RSGplane} it appears clearly delineated through its tangent line-of-sight, pointing towards an associated RSG population.
 
This arm occupies roughly the same space as the so-called molecular ring, a feature visible mostly in molecular emission \citep{Ja06}. Both structures are  mostly young and will be poorly traced by RCGs. As a result, only their tangents are visible in our data and, it is visible in Fig. \ref{ngiants15}, is very difficult to distinguish between the two possible configurations, a closed ring or two tightly wound arms, by looking at both our CMDs and RSG distribution. 

The exact stellar nature of the remaining overdensity in the near end, visible in Fig. \ref{RSGplane} spanning from $l\sim20^\circ$ to $l\sim15^\circ$, has not been established yet. It might be part of the Scutum-Crux arm, might be stellar formation of the bar itself \footnote{Although as stellar formation normally occurs in the dust lanes of bars, hence at these latitudes it should appear beyond the stellar bar itself, our cuts allow for RSGs at these distances to remain in the sample.}, or a part of the molecular ring or the 3-kpc arm\citep{H94} (or any combination of the three). It does not have a clear counterpart at the far end, and this could point to it being part of the 3-kpc arm. As can be seen in Fig. \ref{schem}, the cross section offered by the far pass of the arm to the line-of-sight is much wider, so all the stellar formation that in the first quadrant is visible in $20^\circ\leq l\leq10^\circ$ would appear, due to the effect of projection, around the tangential line-of-sight towards the arm in the fourth quadrant (at $l=-22^\circ$). If there is any stellar formation along the dust lane, in Fig. \ref{RSGplane} it would be visible with $l\leq-13^\circ$ and so it would be mixed with that of the tip of the Long Bar and the increased density towards the Galactic centre. As such, it is almost impossible to distinguish between both hypotheses without a proper spectroscopic study of the individual sources themselves.

\section{Conclusions}
\label{discu}
In this work we obtained combined CMDs of 2MASS and VVV, mapping for the first time the innermost Milky Way at negative galactic longitudes. Using these data along with photometrically selected RSGs -used as proxies for recent stellar formation- we mapped the structures present at these latitudes, offering a fresh picture of the structure of our Galaxy.

The main conclusions drawn follow:
\begin{itemize}
\item[-] Based on the analysis of the CMDs and the 2-D distribution of the RCGs along several lines of sight, we showed there is a flat structure that runs up to $l\sim-12^\circ$. It corresponds to the extension of the Long Bar into the fourth quadrant. With this new constraint, we concluded that it forms an angle of $41^\circ\pm5^\circ$ with the Sun-Galactic centre line.
\item[-] This Long Bar affects the structure of the disc. One of the more severe effects is the truncation of the stellar density with $R_{\mathrm{GC}}\leq5~\mathrm{kpc}$. Because it corresponds with two maxima of the bar potential, this area is devoid of stars and gas (outside the body of the bar itself).
\item[-] Using photometrically selected RSGs as a proxy for recent stellar formation, we demonstrated another of the predicted effects of a galactic bar, the triggering of formation regions at its ends, clearly visible at $l\sim27^\circ$ and $l\sim-12^\circ$. The tangents to the Scutum-Crux and the so-called 3-kpc arm are also visible in the distribution of these stars, showing that in both arms there are traces of recent massive stellar formation.
\end{itemize}
\appendix

\section{VVV+2MASS CMDs of the inner Milky Way}
\label{appdcm}
\begin{figure*}
\includegraphics[width=18.5cm,angle=180]{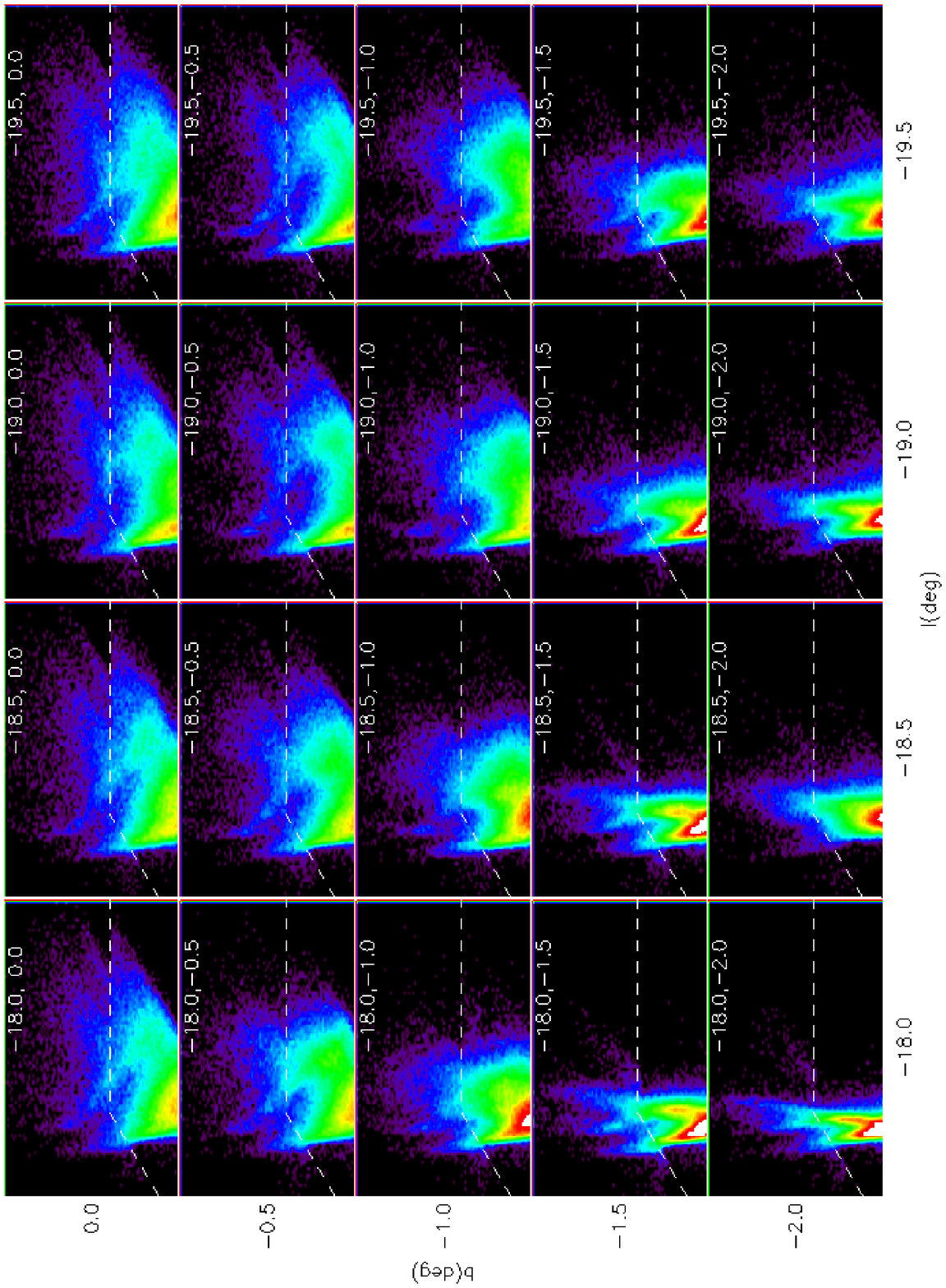}
\caption{CMDs of the in plane fields with $b\leq0^\circ$. The axis and colour scale are those of Fig. \ref{DCM}}
\label{av5}
\end{figure*}
\begin{figure*}
\includegraphics[width=18.5cm,angle=180]{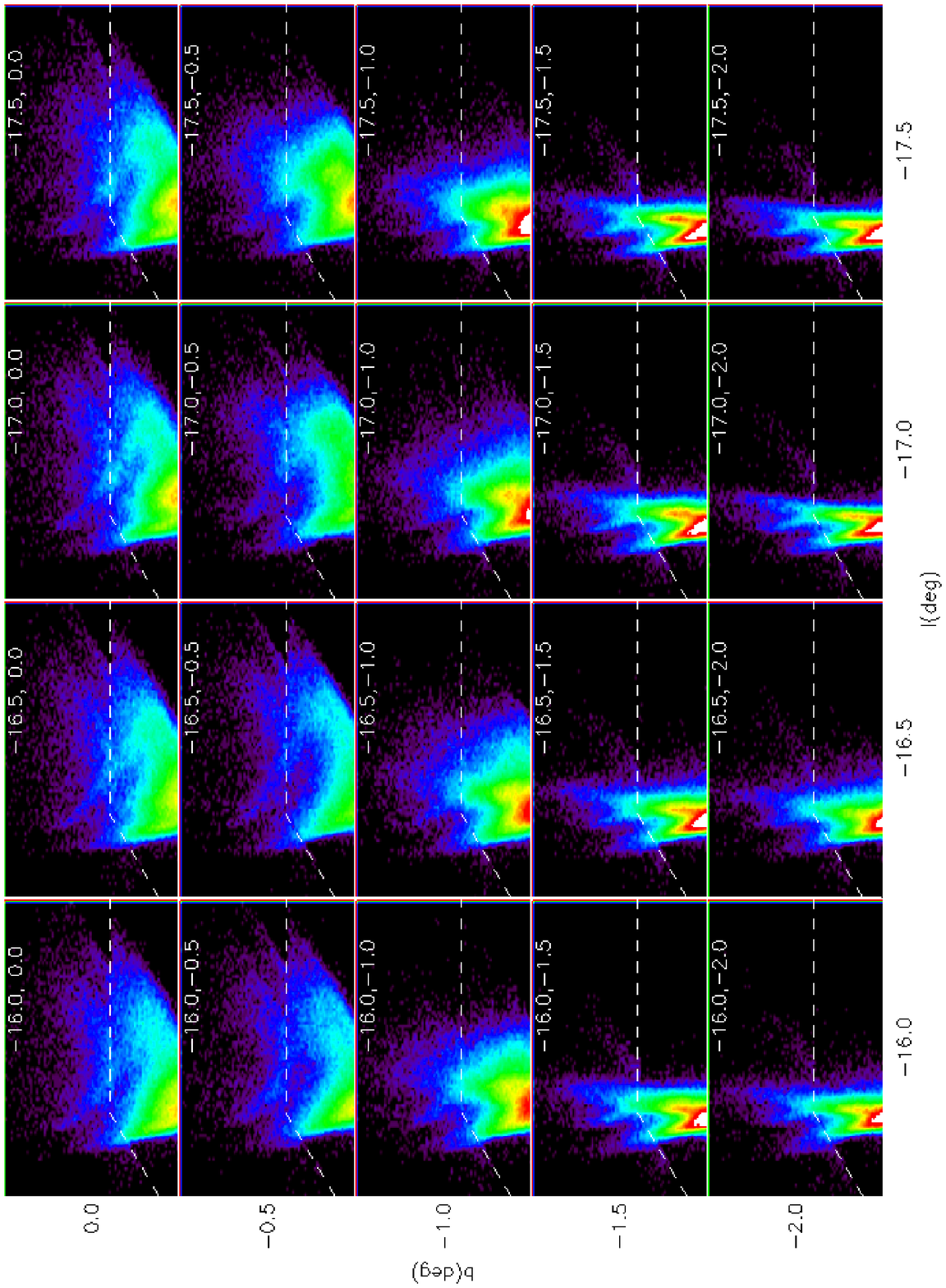}
\caption{CMDs of the in plane fields with $b\leq0^\circ$. The axis and colour scale are those of Fig. \ref{DCM}}
\label{av4}
\end{figure*}
\begin{figure*}
\includegraphics[width=18.5cm,angle=180]{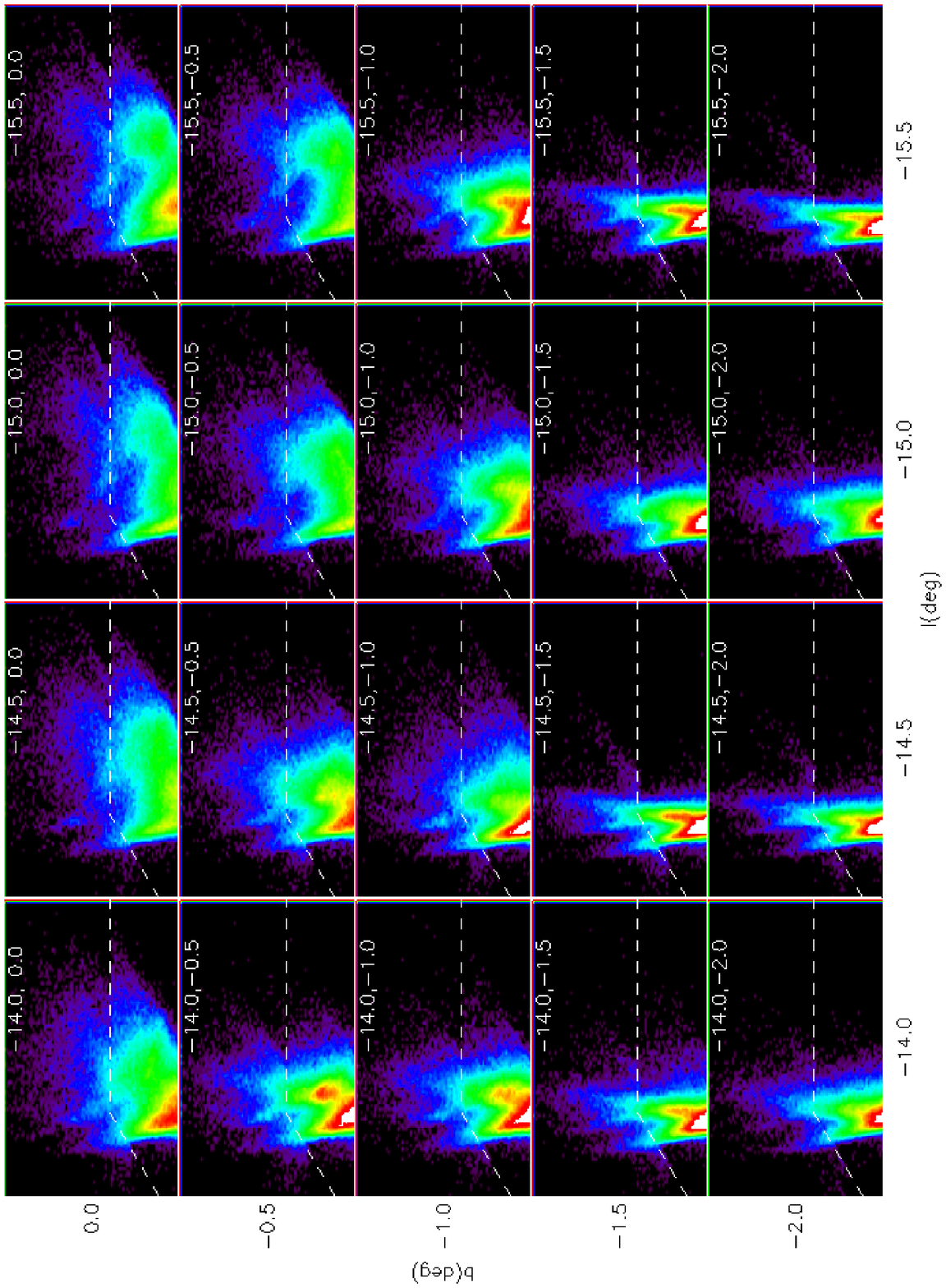}
\caption{CMDs of the in plane fields with $b\leq0^\circ$. The axis and colour scale are those of Fig. \ref{DCM}}
\label{av3}
\end{figure*}
\begin{figure*}
\includegraphics[width=18.5cm,angle=180]{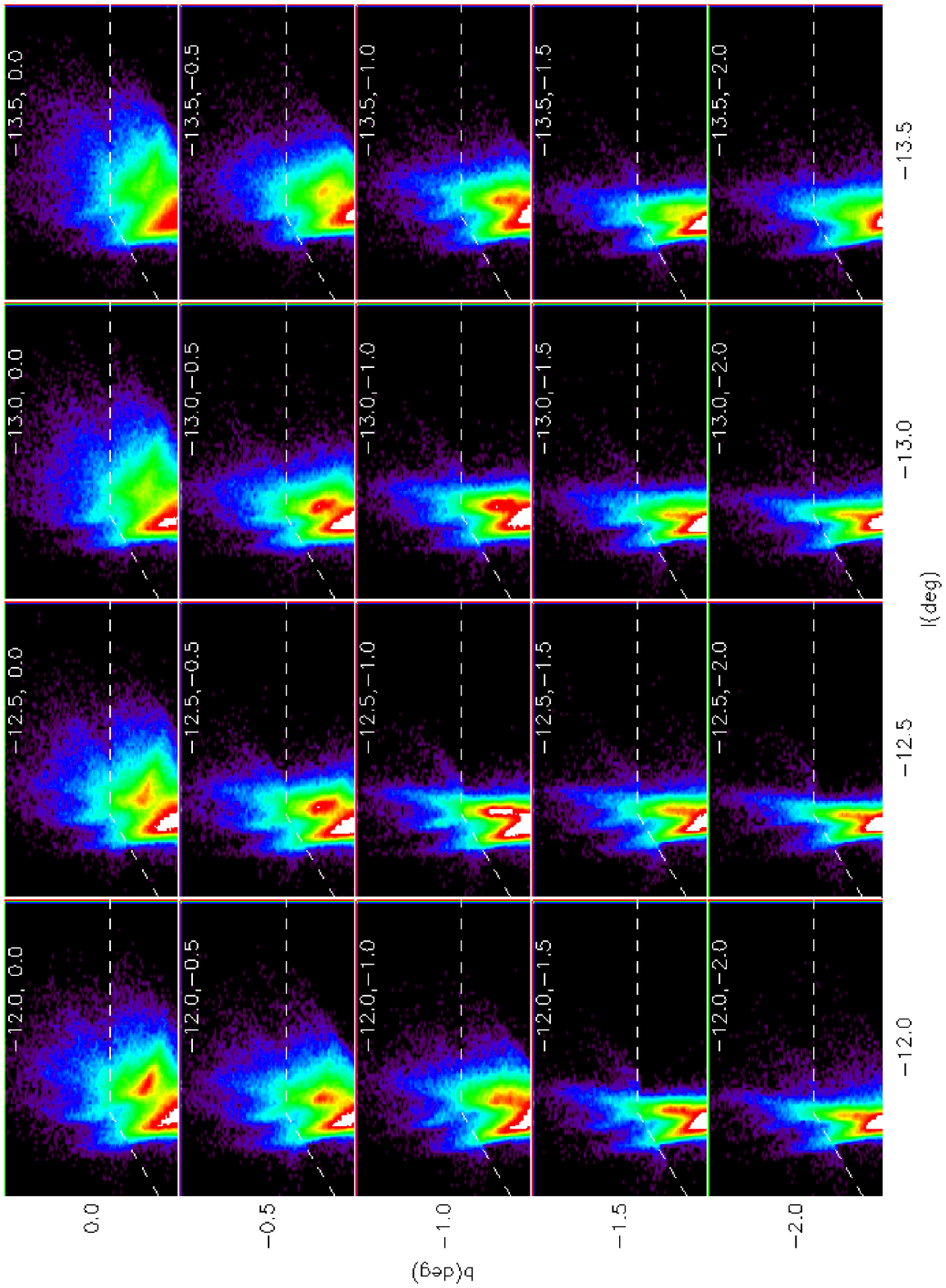}
\caption{CMDs of the in plane fields with $b\leq0^\circ$. The axis and colour scale are those of Fig. \ref{DCM}}
\label{av2}
\end{figure*}
\begin{figure*}
\includegraphics[width=18.5cm,angle=180]{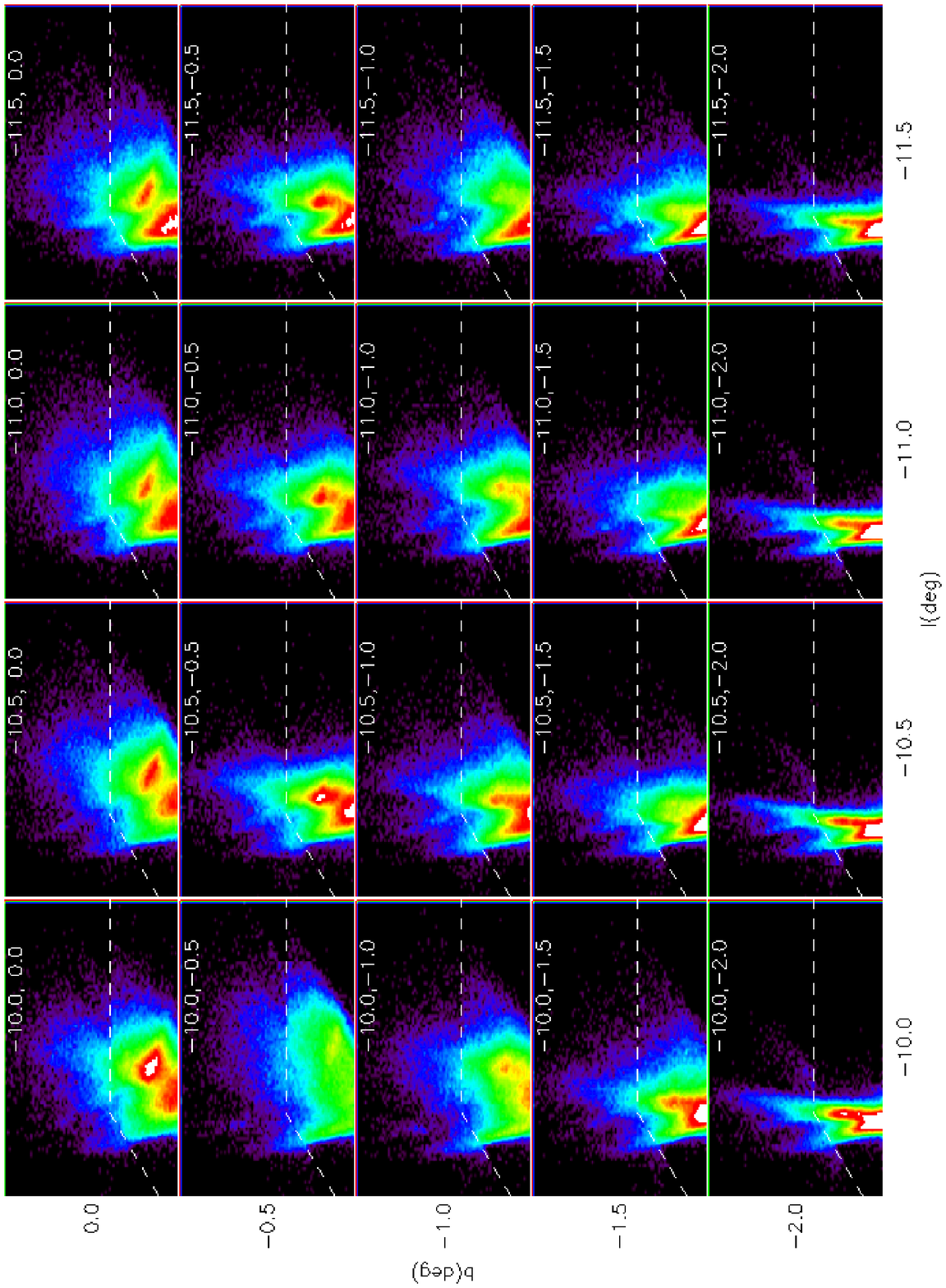}
\caption{CMDs of the in plane fields with $b\leq0^\circ$. The axis and colour scale are those of Fig. \ref{DCM}}
\label{av1}
\end{figure*}
\begin{figure*}
\includegraphics[width=18.5cm,angle=180]{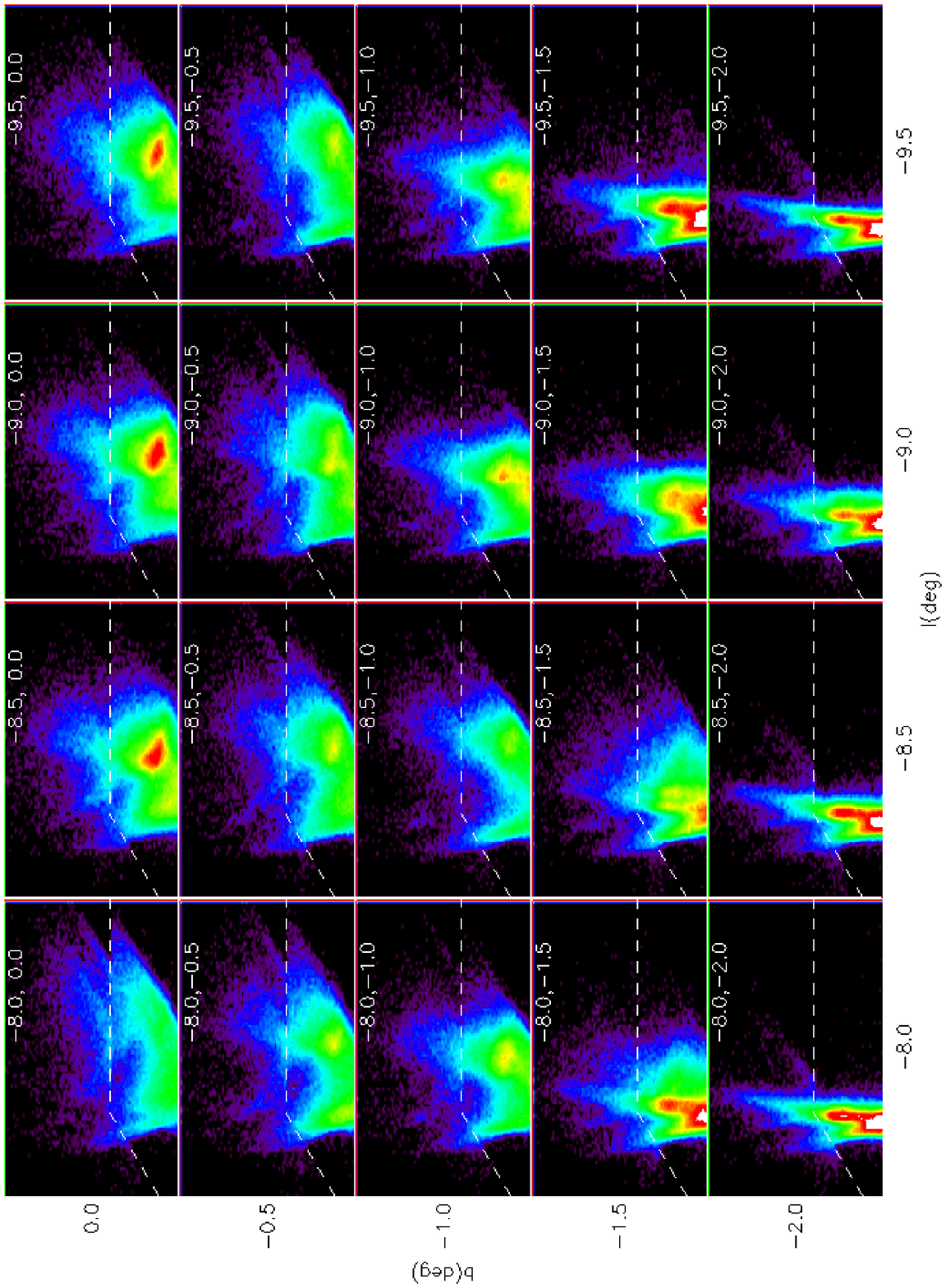}
\caption{CMDs of the in plane fields with $b\leq0^\circ$. The axis and colour scale are those of Fig. \ref{DCM}}
\label{av0}
\end{figure*}

\begin{figure*}
\includegraphics[width=18.5cm,angle=180]{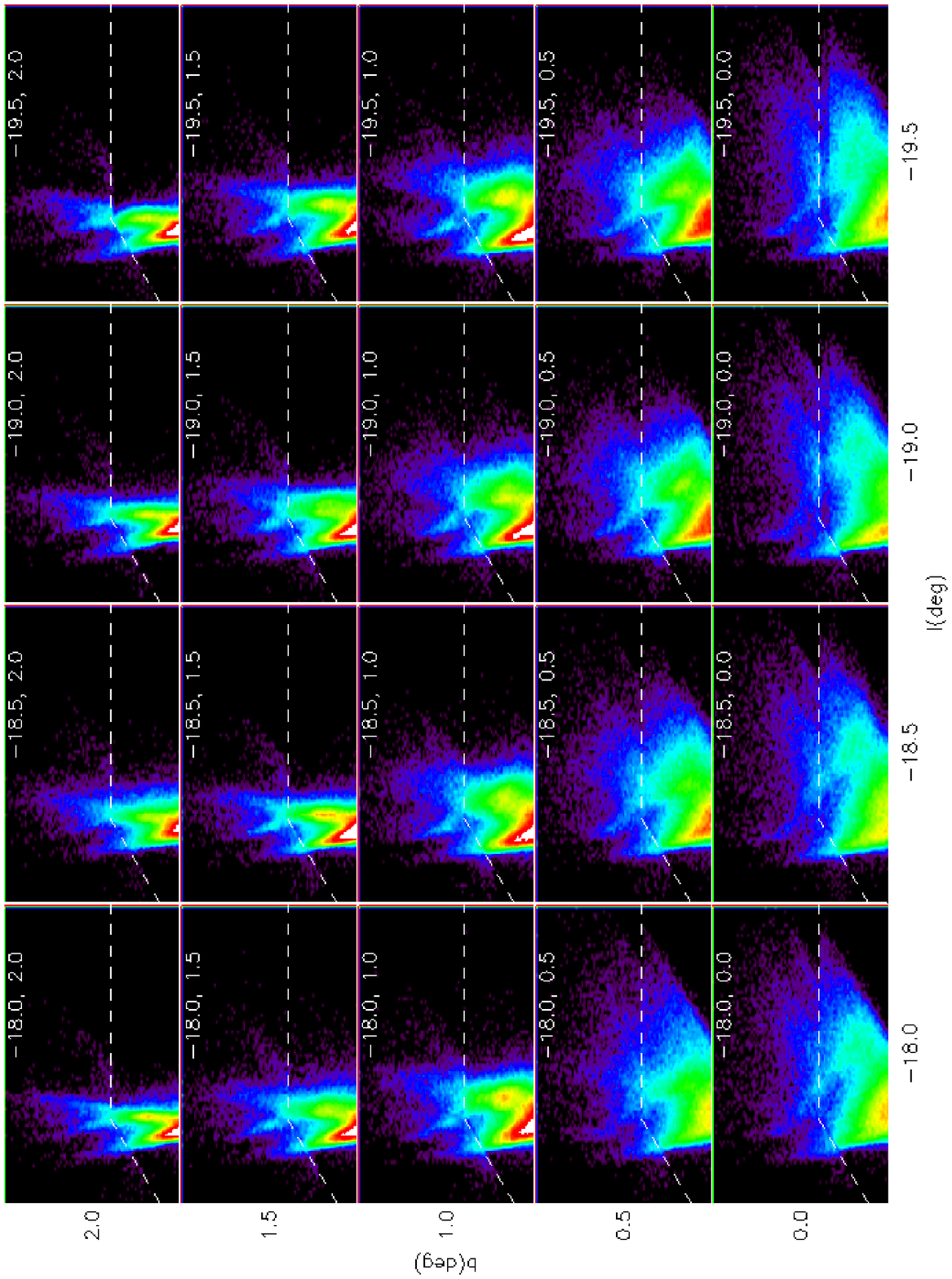}
\caption{CMDs of the in plane fields with $b\geq0^\circ$. The axis and colour scale are those of Fig. \ref{DCM}}
\label{ab5}
\end{figure*}
\begin{figure*}
\includegraphics[width=18.5cm,angle=180]{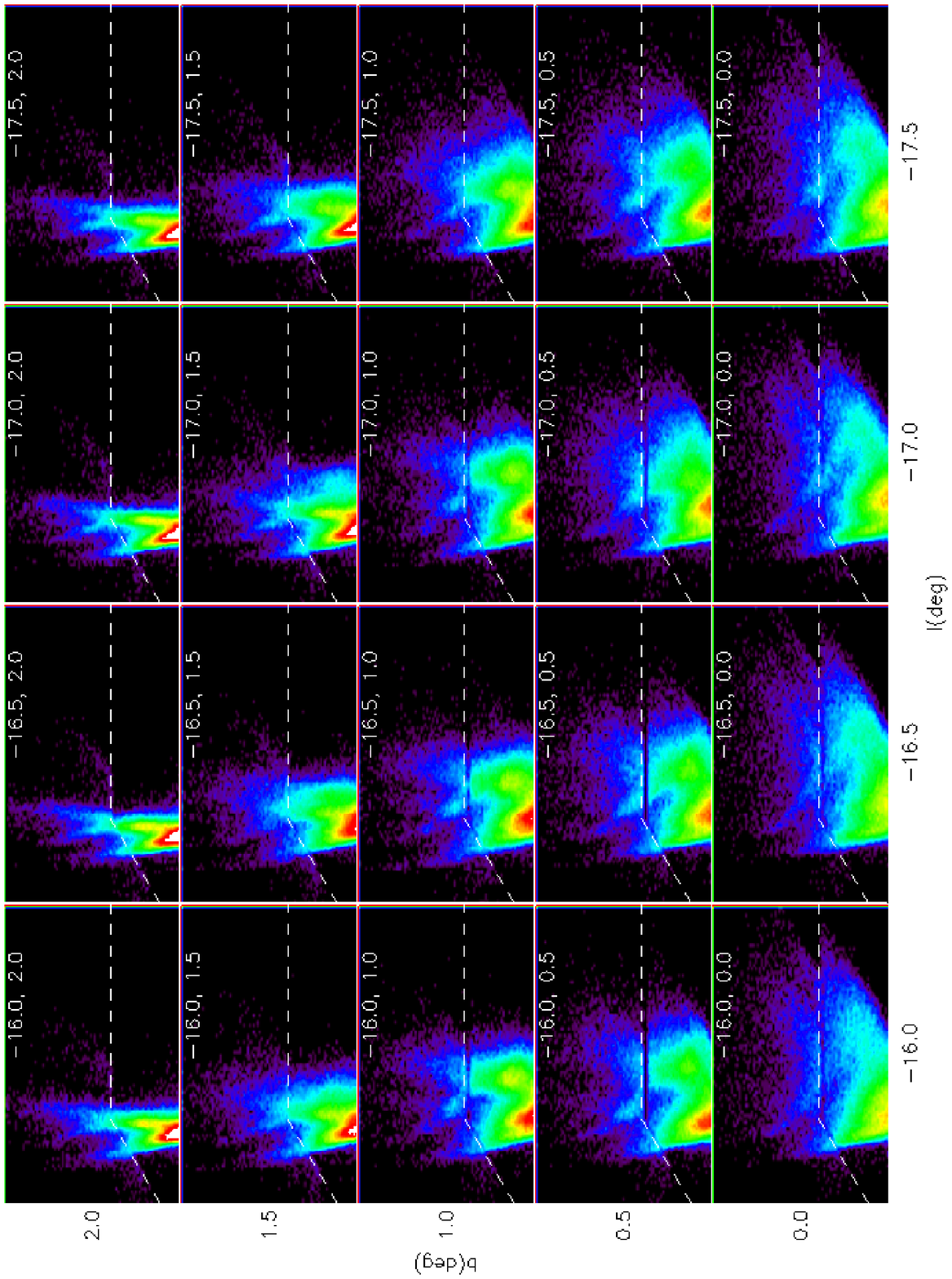}
\caption{CMDs of the in plane fields with $b\geq0^\circ$. The axis and colour scale are those of Fig. \ref{DCM}}
\label{ab4}
\end{figure*}
\begin{figure*}
\includegraphics[width=18.5cm,angle=180]{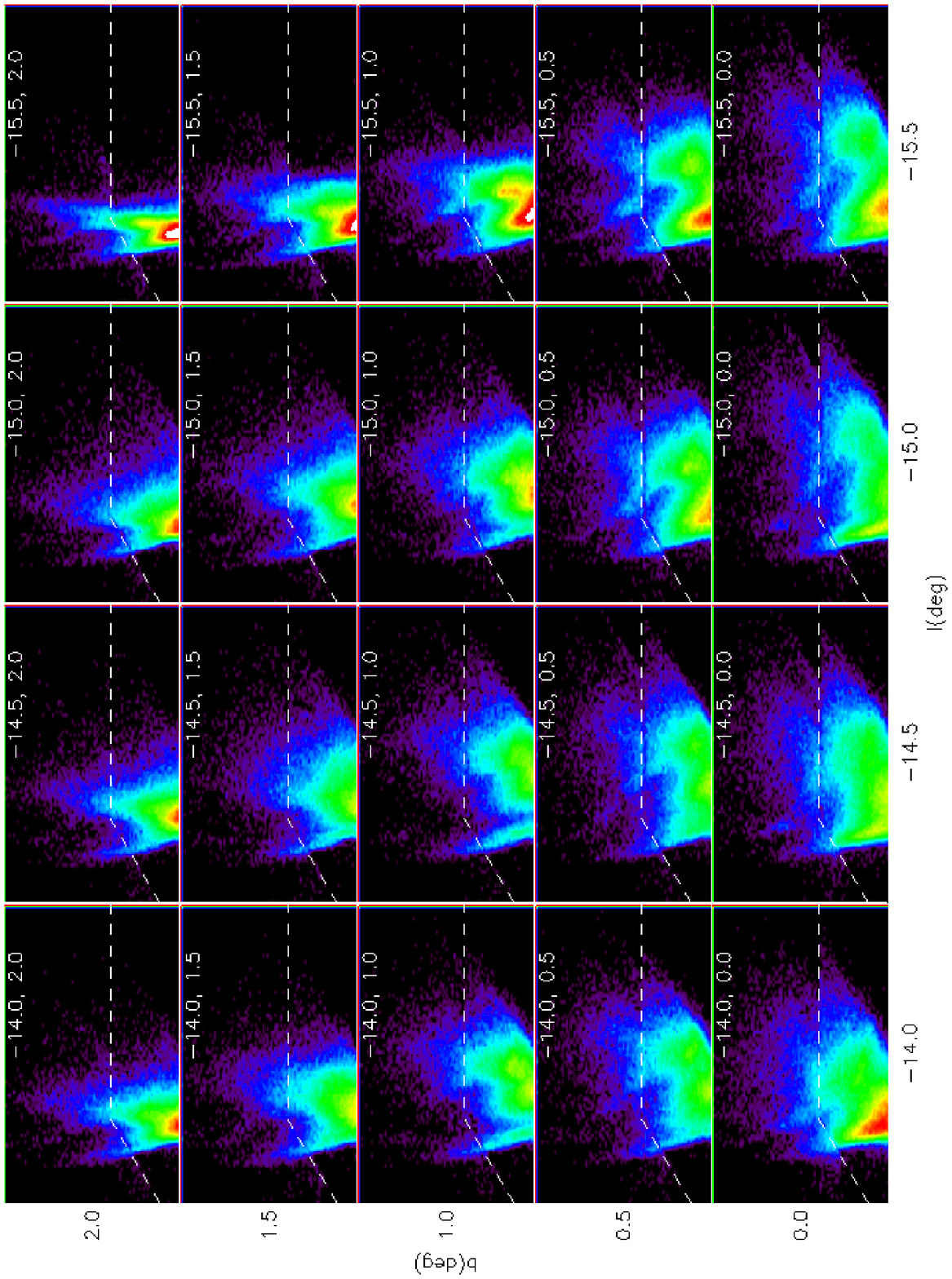}
\caption{CMDs of the in plane fields with $b\geq0^\circ$. The axis and colour scale are those of Fig. \ref{DCM}}
\label{ab3}
\end{figure*}
\begin{figure*}
\includegraphics[width=18.5cm,angle=180]{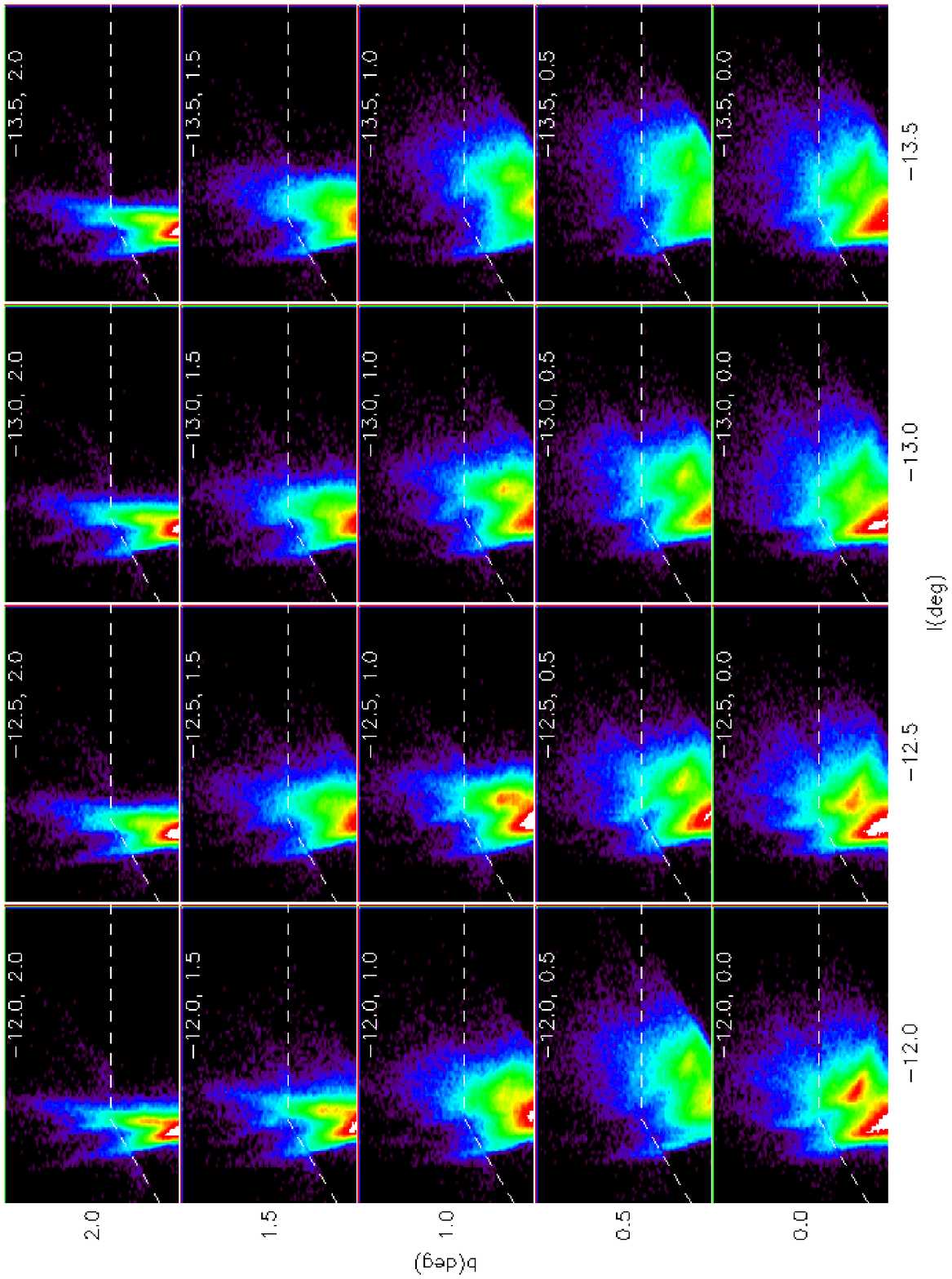}
\caption{CMDs of the in plane fields with $b\geq0^\circ$. The axis and colour scale are those of Fig. \ref{DCM}}
\label{ab2}
\end{figure*}
\begin{figure*}
\includegraphics[width=18.5cm,angle=180]{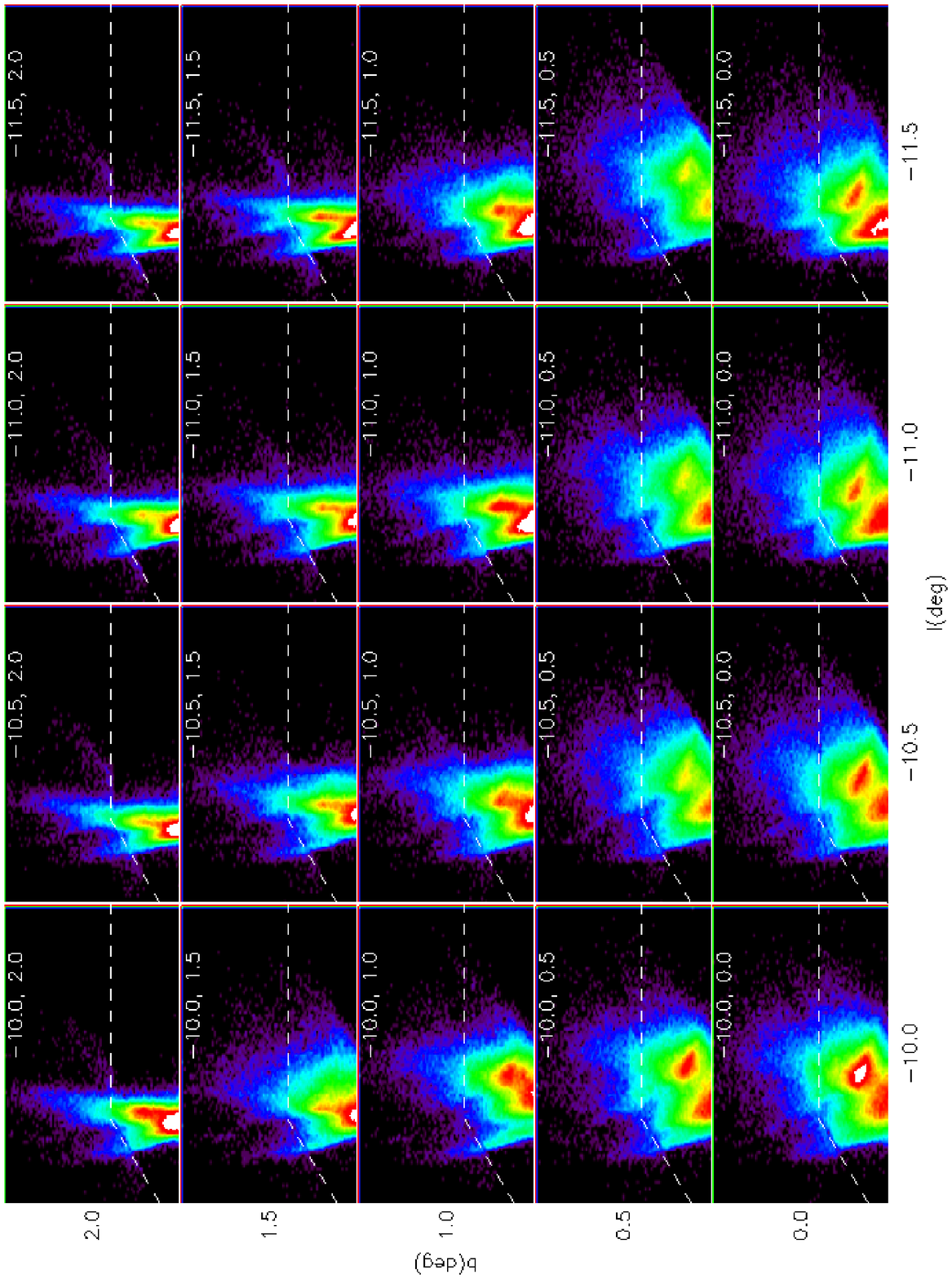}
\caption{CMDs of the in plane fields with $b\geq0^\circ$. The axis and colour scale are those of Fig. \ref{DCM}}
\label{ab1}
\end{figure*}
\begin{figure*}
\includegraphics[width=18.5cm,angle=180]{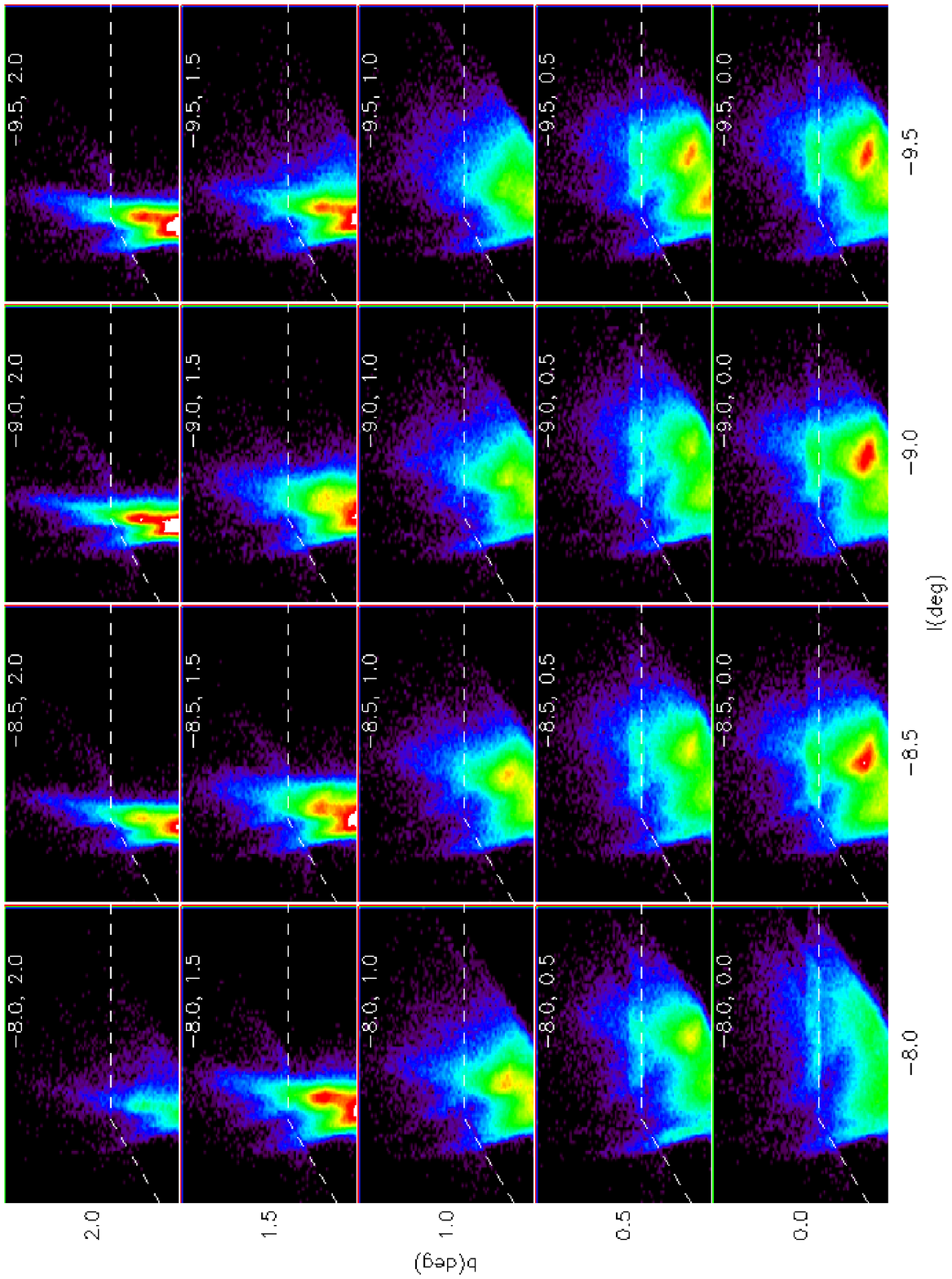}
\caption{CMDs of the in plane fields with $b\geq0^\circ$. The axis and colour scale are those of Fig. \ref{DCM}}
\label{ab0}
\end{figure*}

\begin{acknowledgements}
	  This research is partially supported by the Spanish Ministerio de Ciencia e Innovaci\'on (MCINN) under  AYA2010-21697-C05-5 and the Consolider-Ingenio 2010 Program grant CSD2006-00070: First Science with the GTC (http://www.iac.es/consolider-ingenio-gtc). MLC was supported by the grant AYA2007-67625-CO2-01 of the Spanish Science Ministry. Eduardo Am\^ores obtained financial support for this work from the Funda\c{c}\~{a}o para a Ci\^{e}ncia e Tecnologia (FCT) under the grant SFRH/BPD/42239/2007 and CNPq (311838/2011-1).
      We gratefully acknowledge use of data from the ESO Public Survey programme ID 179.B-2002 taken with the VISTA telescope, data products from the Cambridge Astronomical Survey Unit, and funding from the FONDAP Centre for Astrophysics 15010003, the BASAL CATA Centre for Astrophysics and Associated Technologies PFB-06, the MILENIO Milky Way Millennium Nucleus from the Ministry of Economy's ICM grant P07-021-F, and by Proyecto FONDECYT Regular No. 1090213 from CONICYT.
      This publication makes use of data products from the Two Micron All Sky Survey, which is a joint project of the University of Massachusetts and the Infrared Processing and Analysis Center/California Institute of Technology, funded by the National Aeronautics and Space Administration and the National Science Foundation.
      This research has made use of the NASA/ IPAC Infrared Science Archive, which is operated by the Jet Propulsion Laboratory, California Institute of Technology, under contract with the National Aeronautics and Space Administration. This research has made use of Aladin.
 
\end{acknowledgements}

\bibliographystyle{aa} 
\bibliography{paper.bib} 

\end{document}